\documentclass[12pt,notitlepage,a4paper]{article}
\pdfoutput=1

\usepackage{amssymb}
\usepackage{delarray,amsmath,bbm,epsfig}
\usepackage{cite}
\usepackage{color, graphicx} 

\newcommand\fverb{\setbox\pippobox=\hbox\bgroup\verb}
\newcommand\fverbdo{\egroup\medskip\noindent%
                              \fbox{\unhbox\pippobox}\ }
\newcommand\fverbit{\egroup\item[\fbox{\unhbox\pippobox}]}
\newbox\pippobox
\newcommand{\nn}{\nonumber}
\newcommand{\beq} {\begin{equation}}
\newcommand{\eeq} {\end{equation}}
\newcommand{\beqa} {\begin{eqnarray}}
\newcommand{\eeqa} {\end{eqnarray}}

\newcommand{\eg}{{\it e.g.}}

\newcommand{\eq}[1]{(\ref{eq:#1})}

\newcommand{\inv}[1]{\frac{1}{#1}}

\newcommand{\la}{\lambda}

\newcommand{\re}{{\rm Re}}

\newcommand{\mC}{\mathcal{C}}

\newcommand{\mE}{\mathcal{E}}

\newcommand{\be}{\begin{equation}}
\newcommand{\ee}{\end{equation}}
\newcommand{\bea}{\begin{eqnarray}}
\newcommand{\eea}{\end{eqnarray}}

\def\la{\lambda}

\newcommand{\zbar}{\bar{z}}
\newcommand{\Qbar}{\bar{Q}}

\begin{document}

 \begin{flushright}
 CCTP-2011-40 \\
 HIP-2011-32/TH
 \end{flushright}

 \vskip 2cm \centerline{\Large {\bf New results for the SQCD Hilbert series}} \vskip 1cm
 \renewcommand{\thefootnote}{\fnsymbol{footnote}}
 \centerline{{\bf Niko
 Jokela,$^{1,2}$\footnote{niko.jokela@usc.es} Matti
 J\"arvinen,$^{3}$\footnote{mjarvine@physics.uoc.gr} and Esko
 Keski-Vakkuri$^{4,5,6}$\footnote{esko.keski-vakkuri@helsinki.fi}}}
 \vskip .5cm

 \centerline{\it ${}^1$Departamento de  F\'\i sica de Part\'\i culas}
 \centerline{\it Universidade de Santiago de Compostela}
 \centerline{\it and}
 \centerline{\it ${}^{2}$Instituto Galego de F\'\i sica de Altas Enerx\'\i as (IGFAE)}
 \centerline{\it E-15782, Santiago de Compostela, Spain}
 \centerline{\it{ \ }}

\centerline{\it ${}^3$Crete Center for Theoretical Physics}
\centerline{\it Department of Physics}
\centerline{\it University of Crete, 71003 Heraklion, Greece}

\centerline{\it}
\centerline{\it ${}^{4}$Helsinki Institute of Physics and ${}^{5}$Department of
 Physics } \centerline{\it P.O.Box 64, FIN-00014 University of Helsinki, Finland}
\centerline{\it and}
\centerline{\it
${}^{6}$Department of Physics and Astronomy, Uppsala University}
\centerline{\it SE-75108 Uppsala, Sweden}

\setcounter{footnote}{0}
\renewcommand{\thefootnote}{\arabic{footnote}}

 \begin{abstract}
 We derive new explicit results for the Hilbert series of ${\cal N}=1$ supersymmetric QCD with $U(N_c)$
 and $SU(N_c)$ color symmetry. We use two methods which have previously been applied to similar computational
 problems in the analysis of decay of unstable D-branes: expansions using Schur polynomials, and the log-gas
 approach related to random matrix theory.
 \end{abstract}

 \newpage

\tableofcontents


\section{Introduction and summary}

Hilbert series of a supersymmetric gauge theory is a partition function that contains information
about the spectrum of operators and the moduli space of vacuum solutions, and has been used to study many properties
of a large class of theories. An introduction to and review of the subject is, {\em{e.g.}}, the PhD thesis \cite{Forcella:2009bv}.
In this paper we focus on a narrowly defined calculational problem: computing explicit results
for the Hilbert series of ${\cal N}=1$ supersymmetric
QCD with $U(N_c)$ or $SU(N_c)$ color symmetry and $N_f$ flavors in the fundamental representation of $U(N_f)$
flavor symmetry. The standard method to calculate the Hilbert series is based on the plethystic
programme \cite{BFHH,Butti:2006au,Noma:2006pe,Hanany:2006uc,feng,forcella,Butti:2007jv,Forcella:2007ps,hanany,Balasubramanian:2007hu,Forcella:2008bb,Gray:2008yu} and the
Molien-Weyl formula \cite{Gray:2008yu,Pouliot:1998yv,Romelsberger:2005eg}.
This method typically leads to expressions with a large number of contour integrals, which makes it difficult
to obtain explicit exact results or approximations for the Hilbert series. Recently, an alternative method
was presented in \cite{CM} (see also \cite{Basor:2011da} for the $SO(N_c)$ and $Sp(N_c)$ symmetries), based on rewriting the Hilbert series as a basic object of interest in random matrix
theory -- a determinant of a Toeplitz matrix of Fourier coefficients of a periodic function, or in short a
Toeplitz determinant. In random matrix theory they result from the problem of computing expectation values
in an ensemble of unitary random matrices. With this method, a new set of tools, developed
in recent years in random matrix theory, become available and yield explicit results for the Hilbert series
of SQCD.

In this paper we discuss two additional techniques and calculate with them new explicit results for the Hilbert
series. Our motivation comes from previous work where we have been applying and developing the techniques in a different context,
which we now review briefly. Similar random matrix theory integrals or Toeplitz determinants as in \cite{CM} appear in the problem of
computing amplitudes for the decay of unstable D-branes into various closed string
or open string channels. In that context matrix integrals appear in an approach involving correlation functions
in timelike boundary Liouville theory (TBL) \cite{Gutperle:2003xf}, a scalar field theory with the action
\be
  S = -\frac{1}{2\pi} \int_{\Sigma} dzd\zbar~\partial \phi \bar{\partial} \phi
  + \lambda \int_{\partial \Sigma} d\tau~e^{\phi (\tau)} \ ,
\ee
where the last term is a deformation by a Liouville term inserted at the boundary of the worldsheet, and the kinetic
term of the scalar has a negative sign. A straightforward path to compute correlation functions is to treat
the boundary deformation term as a perturbation and expand it as a series in the path integral. After performing the contractions, the $N$th term
in the perturbation series can be recognized as a $U(N)$ matrix integral, with the standard Haar measure coming
from the contractions among the exponential terms \cite{LNT,VEPA,VEPA2}.\footnote{Incidentally, as a precursor of connections
to Yang-Mills theory, the appearance of $U(N)$ matrix integrals inspired \cite{CL} to speculate that TBL is related
to QCD in two dimensions.} For explicit results for the correlation functions, one needs to compute the $U(N)$ integrals. This becomes more
difficult for $n$-point correlation functions with $n\geq 2$ \cite{npt,vikareview}, and involves similar integrals as in the Hilbert series problem:
\bea \label{eq:Zndef}
 Z_{n}(\{w_i \};N) & = & \frac{1}{N!} 
\int\prod_{a=1}^N\frac{d\tau_a}{2\pi}\prod_{1\leq a<b\leq N}|e^{i\tau_a}-e^{i\tau_b}|^2\prod_{a=1}^N\prod_{i=1}^n|1-w_i e^{-i\tau_a}|^{2\xi_i} \ ,
\eea
where $w_i$ are $n$ points in a unit disk in complex plane. We will review the Hilbert series and its connection to
(\ref{eq:Zndef}) in Section \ref{sec:schur}.

We discuss two techniques that have been applied to the integrals (\ref{eq:Zndef}), and will hence also apply to the
Hilbert series: i) the use of orthogonal polynomials on the unit circle (Schur polynomials), and ii) relating the integrals to partition
functions of log-gases with external charges, and then using electrostatics
for saddle point estimates. 
Schur polynomials have appeared in the context of Hilbert series as
the characteristics of $SU(N)$
representations \cite{Gray:2008yu,CM} and they have also been used extensively in the 
analysis of the $\mathcal{N}=4$ super Yang-Mills \cite{Dolan:2007rq}.\footnote{It is interesting to note that
recent studies of $\mathcal{N}=2$ dualities utilize Macdonald polynomials,
which are generalizations of Schur polynomials \cite{varta1,Gadde:2011uv}. Moreover, in some cases 
the superconformal index can be reduced to the Hilbert series \cite{varta2}.}
In this paper we derive several new results for the Hilbert series based on standard
properties of Schur polynomials.
The second technique to be discussed in this paper, the log-gas approach,
has been studied by the authors previously
in \cite{bulk2pt,vikareview} as a way to obtain useful approximations to integrals (\ref{eq:Zndef}).

The paper is organized as follows. We begin by very briefly introducing the refined and unrefined Hilbert series, followed by a brief review of Schur
polynomials. Then, using Schur polynomials, we derive an explicit formula for the refined
Hilbert series when $N_f\leq N_c$. 
The formula is a summed simplification of a series
expression conjectured in \cite{CM}.
As a special case,
the formula reduces to the result of \cite{CM} for the unrefined series. When $N_f > N_c$, for the refined series we
obtain an expression in terms of sums of Young tableaux.
We also derive explicit results in a few cases with small $N_f-N_c$.
We then analyze the singularity structure of the (unrefined) Hilbert series.
A direct calculation verifies that the orders of the poles of the Hilbert series agree with the
dimensions of the moduli space of SQCD also for $N_f>N_c$, where only a few explicit results are known. For a general discussion on the classical and quantum moduli
spaces of vacua of supersymmetric gauge theories, see \cite{Seiberg:1994bz}.

We then move to the second approach, the use of (electrostatics of) log-gas of point particles confined to the unit circle (also called the Dyson gas \cite{Dyson:1962es}).
We rewrite the unrefined Hilbert series as a canonical partition function of a log-gas (with the number of particles depending on $N_c$)
with an external charge (with the charge depending on $N_f$). This allows us to observe a physical phenomenon that is associated with the $N_f=N_c$ threshold.
In the Veneziano limit of large $N_f,N_c$ with fixed ratio $N_f/N_c$, it is possible to obtain an asymptotic approximation for the Hilbert series, by approximating the log-gas of charges with
a continuous charge distribution. The leading contribution comes from the electrostatic minimum
energy configuration for the continuous charge distribution interacting with the external charge.
When $N_f<N_c$, the continuous distribution has nonzero density everywhere, and the leading approximation
for the Hilbert series agrees with the result from the Schur polynomial method. When $N_f>N_c$, exact
methods to compute the series lead to more complicated results.
In the Veneziano limit, in the electrostatic
method we observe that a gap created by the external charge may appear in the continuous charge distribution.\footnote{A similar observation has been made in mathematics literature
\cite{BNR}. Another, earlier related result is \cite{Mandal} which studied the phase structure
of unitary matrix models with polynomial potentials. A difference is that in \cite{Mandal} the electric
fields which arise are not created by point-like external charges.} This is reminiscent of \cite{grosswitten} and we also find a third order phase transition.
The asymptotic result for the Hilbert series is still straightforward to
find, and it again is simple to read off the dimension of the moduli space of 
SQCD.
We also compare our approach to that of \cite{CM}
based on the Toeplitz determinants and evaluating them with the Geronimo-Case-Borodin-Okounkov (GCBO) formula.
We show that this method works in the region where no gap appears in the asymptotic charge distribution of the log-gas method. 

Let us summarize in the following the main new results of this paper:
\begin{itemize}
\item Color group $U(N_c)$
\begin{itemize}
 \item  We prove the general expression (\ref{eq:Ungen}) for the refined $U(N_c)$ Hilbert series as the sum over semi-simple Young tableaux in Section \ref{sec:resultsforUNc}.
 \item We obtain explicit formulas for the refined $U(N_c)$ Hilbert series for $N_f\leq N_c$, $N_f=N_c+1$, and $N_f=N_c+2$, in Eqs, (\ref{eq:unrefUN}),~(\ref{eq:Ncplus1res}), and~(\ref{eq:Nfp2res}), respectively.
 \item For the unrefined series, 
we work out the order of the (only) non-trivial pole in Eq.~\eqref{eq:UNasympt} 
which matches the dimension of the moduli space.
 \item We calculate the unrefined Hilbert series, and the residue of its pole, in the Veneziano limit up to corrections suppressed by $1/N_c^2$ for all reasonable values of the parameters. Results are given in Eqs.~\eqref{eq:lead1},~\eqref{eq:leadgap}, and~\eqref{eq:lead2}.
\end{itemize}
\item Color group $SU(N_c)$
\begin{itemize}
 \item  We prove the general expression  (\ref{eq:Ik}) for the refined $SU(N_c)$ Hilbert series as the sum over semi-simple Young tableaux in Section \ref{sec:resultsforSUNc}.
 \item We obtain explicit formulas for the refined $SU(N_c)$ Hilbert series for $N_f< N_c$, $N_f=N_c$, and $N_f=N_c+1$, in Eqs, (\ref{eq:SUNcNflessthanNc}),~(\ref{eq:SUNcNfequalsNc}), and~(\ref{eq:SUNcp1res}), respectively.
 \item We work out the orders of most of the poles of the unrefined series in Eqs.~\eqref{eq:SUt1sing} and~\eqref{eq:SUtt1sing}, and find again agreement with the dimension of the moduli space.
 \item We prove that the results for the unrefined $U(N_c)$ Hilbert series in the Veneziano limit also apply to the $SU(N_c)$ Hilbert series, with certain restrictions on the values of its parameters (see Eq.~\eqref{eq:USUres}).
\end{itemize}
\end{itemize}

\section{Hilbert series and Schur polynomials} \label{sec:schur}

\subsection{Hilbert series}

As in \cite{CM}, we begin by considering an ${\cal N}=1$ supersymmetric $U(N_c)$ Yang-Mills theory with $N_f$ flavors of quarks
and anti-quarks in the fundamental and antifundamental representation. The theory has various global symmetries; of particular interest
here are the $U(1)_Q$ ($U(1)_{\Qbar}$) charge symmetries, with quarks (anti-quarks) carrying charge
$+1$ $(-1)$, and the $SU(N_f)_1\times SU(N_f)_2$ flavor symmetry. The refined Hilbert series is a partition function
\cite{Forcella:2009bv,hanany} involving fugacities associated with the color and flavor indices and with the global
charges. We denote the $SU(N_f)_1\times SU(N_f)_2$ flavor fugacities by $x_i,y_i$, with $i=1,\ldots ,N_f$ and the $U(N_c)$ color
fugacities by $z_a= e^{i\tau_a}$, with $a=1,\ldots, N_c$.
The $U(1)_Q\times U(1)_{\Qbar}$ fugacities are denoted by
$t, \tilde t$. With the help of the Molien-Weyl formula from representation theory, the refined Hilbert series can
be written in a simple-looking form which however involves nested integrals. After rescaling the flavor fugacities, introducing
\begin{eqnarray}
 \tilde x = \left(\tilde x_1,\tilde x_2,\ldots,\tilde x_{N_f}\right) = \left(x_1,\frac{ x_2}{x_1},\frac{ x_3}{x_2},\ldots,\frac{1}{x_{N_f-1}}\right) \\
 \tilde y = \left(\tilde y_1,\tilde y_2,\ldots,\tilde y_{N_f}\right) = \left(\frac{1}{y_1},\frac{ y_1}{y_2},\frac{ y_2}{y_3},\ldots,y_{N_f-1}\right) \  \ ,
\end{eqnarray}
the refined Hilbert series becomes a Selberg type integral
\begin{equation} \label{eq:refdef}
 g_{N_f,U(N_c)}(t,\tilde t,x,y) = \frac{1}{N_c!}\prod_{a=1}^{N_c}  \int_0^{2\pi}\frac{d\tau_a}{2\pi}|\Delta(z)|^2 \prod_{i=1}^{N_f}\prod_{a=1}^{N_c}\frac{1}{\left(1-t \tilde x_i z_a^{-1}\right)\left(1- \tilde t \tilde y_i z_a\right)} \ ,
\end{equation}
where $\Delta(z)$ is the Vandermonde determinant
\begin{equation}
 \Delta(z)= \prod_{1\le a < b \le N_c }(z_a-z_b) \ .
\end{equation}
Notice that
\begin{equation} \label{eq:xprodid}
 \prod_{i=1}^{N_f} \tilde x_i = 1 =  \prod_{i=1}^{N_f} \tilde y_i \ .
\end{equation}

For the reduction to $SU(N_c)$ color symmetry, we introduce the extra constraint that the phases $\tau_a$ sum to zero (modulo $2\pi$):
\begin{eqnarray} \label{eq:refdefSU}
 g_{N_f,SU(N_c)}(t,\tilde t,x,y) &=& \frac{1}{N_c!}\prod_{a=1}^{N_c}  \int_0^{2\pi}\frac{d\tau_a}{2\pi}\sum_{k=-\infty}^\infty 2\pi\delta\left(\sum_a\tau_a-2\pi k\right) \  |\Delta(z)|^2 \nn\\
&&\ \  \times \prod_{i=1}^{N_f}\prod_{a=1}^{N_c}\frac{1}{\left(1-t \tilde x_i z_a^{-1}\right)\left(1- \tilde t \tilde y_i z_a\right)} \ .
\end{eqnarray}
As pointed out in \cite{CM},
this result  can be written as a sum over unconstrained integrals as
\begin{eqnarray} \label{eq:SULagrexp}
  g_{N_f,SU(N_c)}(t,\tilde t,x,y) & = &\sum_{k=-\infty}^\infty I_k(t,\tilde t,x,y) \nonumber\\
I_k(t,\tilde t,x,y) &\equiv& \frac{1}{N_c!}\prod_{a=1}^{N_c}  \int_0^{2\pi}\frac{d\tau_a}{2\pi}|\Delta(z)|^2 z_1^kz_2^k\cdots z_{N_c}^k \nonumber\\
 &  & \ \ \ \ \times \prod_{i=1}^{N_f}\prod_{a=1}^{N_c}\frac{1}{\left(1-t \tilde x_i z_a^{-1}\right)\left(1-\tilde t \tilde y_i z_a\right)} \ ,
\end{eqnarray}
where $I_0$ equals the $U(N_c)$ partition function $g_{N_f,U(N_c)}$.

The unrefined Hilbert series are obtained by the setting $x_1=\cdots =x_{N_f}=y_1=\cdots =y_{N_f}=1$ in the above definitions, and we shall denote them by $g_{N_f,U(N_c)}(t,\tilde t)$ and by  $g_{N_f,SU(N_c)}(t,\tilde t)$ for the $U(N_c)$ and $SU(N_c)$ gauge groups, respectively.

It is well known that the matrix integrals~\eqref{eq:refdef} and~\eqref{eq:SULagrexp} can be rewritten as $N_c \times N_c$ Toeplitz determinants \cite{CM}. 
For the case of unrefined series, the elements of the determinants can be evaluated explicitly (see Appendix~\ref{AppToeplitz}), leading to explicit expressions for the unrefined series for small $N_c$ and arbitrary $N_f$. While Toeplitz determinants can also be defined for the refined series, it is more efficient to use the language of Schur polynomials, which we shall discuss next.

\subsection{Schur polynomials}

The integral \eq{refdef} has been previously analyzed in the context of D-brane decay \cite{CL,bulk2pt}. The method presented in \cite{CL} uses
expansions in a particular set of symmetric polynomials called the Schur polynomials, and applies directly to the present case. For completeness,
we shall review the calculation here.

To begin with, let $\lambda =(\lambda_1,\lambda_2,\ldots,\lambda_n)$, be a partition
of  $|\lambda|=\sum_i\lambda_i$ with $\lambda_1 \ge \lambda_2 \ge \cdots \ge \lambda_n \ge 0$.
Then $\lambda$ parametrizes a Young tableau, corresponding to a Young diagram where the $i$th column
contains $\lambda_i$ boxes. We also define the length of the tableau $\ell(\lambda)=\max\{i:\lambda_i>0\}$.
The Schur polynomials of the variables $z=(z_1,z_2,\ldots,z_n)$ are characterized by $\lambda$:
\begin{equation} \label{eq:schur}
 s_\lambda(z)= s_\lambda(z_1,z_2,\ldots,z_n) = \frac{\det\left(z_j^{\lambda_{n-i+1}+i-1}\right)_{i,j=1,\ldots,n}}{\det\left(z_j^{i-1}\right)_{i,j=1,\ldots,n}} = \frac{\det\left(z_j^{\lambda_{n-i+1}+i-1}\right)_{i,j=1,\ldots,n}}{\Delta(z)} \ ,
\end{equation}
where the denominator is the Vandermonde determinant. It is not difficult to see that this expression indeed defines a polynomial which is symmetric under the interchange of any two variables.

We shall need the extension for the definition of the Schur polynomials to the cases where the length of the partition exceeds the number of variables, $\lambda =(\lambda_1,\lambda_2,\ldots,\lambda_m)$ with $m>n$. It is convenient to define $s_\la(z)=0$ whenever $\ell(\la)>n$, {\em{i.e.}}, the Young diagram is larger than the number of variables, whereas the above definition~\eqref{eq:schur} applies for $\ell(\la) \leq n$. With this extension, we have $s_\la(z_1,z_2,...,z_n,0)=s_\la(z_1,z_2,...,z_n)$.

The polynomials are orthogonal with respect to the Haar measure of \eq{refdef}:
\begin{equation}
\label{eq:orth}
 \frac{1}{n!}\prod_{i=1}^{n} \int_0^{2\pi} \frac{d\tau_i}{2\pi}|\Delta(z)|^2 s_\lambda(z_1,\ldots,z_n) s_\kappa(\bar z_1,\ldots,\bar z_n) =   \prod_{k=1}^n\delta_{\lambda_{k},\kappa_{k}} \equiv \delta_{\lambda,\kappa} \ , 
\end{equation}
as can be verified by direct computation using the definition~\eqref{eq:schur}.\footnote{The orthogonality property holds for $\ell(\lambda),\ell(\kappa) \leq n$. If the length of either of the Young diagrams exceeds $n$, the integrand vanishes.}

The most important tool for us is the Cauchy identity. Let $z=(z_1,\ldots,z_n)$ and $w=(w_1,\ldots,w_m)$. Then
\begin{equation}
\label{eq:Cauchy}
 \prod_{i=1}^n \prod_{j=1}^m \frac{1}{1-z_iw_j} = \sum_\la s_\la(z) s_\la(w) \ .
\end{equation}
Here the sum over $\la$ is restricted by the requirement that both $s_\la(z)$ and $s_\la(w)$ are nonzero, which leads
to $\ell(\la)\le \min(n,m)$.\footnote{Notice that if, \eg, $\ell(\la)<n$, we take $\la_i=0$ for $n\ge i>\ell(\la)$ in
the definition of the Schur polynomial $s_\la(z)$.}

The following property of the Schur polynomials, which follows easily from the definition~\eqref{eq:schur}, will also be useful. For $k\ge 0$, we have
\beq \label{eq:shiftid}
 z_1^kz_2^k\cdots z_{n}^k s_\kappa(z) = s_{\kappa+k}(z)  \qquad (k\ge 0) \ ,
\eeq
where $\kappa+k$ is defined by $(\kappa+k)_i=\kappa_i+k$  with $i=1,\ldots,n$. We shall also denote $\kappa+k =\kappa+k(n)$ to stress that $k$ boxes were added only to the first $n$ columns of the Young diagram whenever this is not clear from the context.

\subsection{Results for the refined $U(N_c)$ series}\label{sec:resultsforUNc}

Let us then apply these results to the refined Hilbert series of Eq.~\eq{refdef}. Applying the Cauchy identity (with $(n,m)=(N_f,N_c)$) we have
\begin{equation}
 \prod_{i=1}^{N_f}\prod_{a=1}^{N_c}\frac{1}{\left(1-t \tilde x_i z_a^{-1}\right)\left(1-t \tilde y_k z_a\right)} = \sum_{\la,\kappa} t^{|\la|} s_\la(\tilde x) s_\la(\bar z)\tilde  t^{|\kappa|} s_\kappa(\tilde y) s_\kappa(z) \ ,
\end{equation}
where $\ell(\la),\ell(\kappa) \le \min(N_f,N_c)$ and we used $s_\la(tz_1,\ldots,tz_{N_f}) = t^{|\la|}s_\la(z_1,\ldots,z_{N_f})$. Using the orthogonality relation of Eq.~\eq{orth}, we get
\begin{eqnarray} \label{eq:Ungen}
  g_{N_f,U(N_c)}(t,\tilde t,x,y) &=&  \sum_{\la,\kappa} t^{|\la|} s_\la(\tilde x) \tilde t^{|\kappa|} s_\kappa(\tilde y) \delta_{\la,\kappa} \nonumber \\
&=&\sum_{\la:\ell(\la) \le \min(N_f,N_c)} (t\tilde t)^{|\la|} s_\la(\tilde x)  s_\la(\tilde y) \ , \
\end{eqnarray}
for all $N_f,N_c$.
If $N_f\le N_c$, the constraint $\ell(\la) \le \min(N_f,N_c)=N_f$ is automatically satisfied. In this case we can use the Cauchy identity again, and get
\begin{equation}
\label{eq:unrefUN}
 g_{N_f,U(N_c)}(t,\tilde t,x,y) = \prod_{i=1}^{N_f} \prod_{j=1}^{N_f}\frac{1}{1-t\tilde t \tilde x_i \tilde y_j}
  \  \qquad (N_f \le N_c)
\end{equation}
which is independent of $N_c$. 

Ref. \cite{CM} conjectured the formula~\eqref{eq:Ungen} for the refined Hilbert series, in the form of an infinite series
with coefficients given by $SU(N_f)$ characters (eqn. (2.79) in \cite{CM}). We have rewritten the formula in terms of the Schur polynomials, and proven it here. 
For $N_f \leq N_c$, the infinite series expansion then becomes equal to the right hand
side of the Cauchy identity (\ref{eq:Cauchy}), so that the expansion reduces to the form (\ref{eq:unrefUN})
above.
This explicit result for the refined Hilbert series is manifestly
a generalization of Eq. (2.60) in \cite{CM} for the unrefined series,
which was obtained using the Toeplitz determinant method and the Geronimo-Case-Borodin-Okounkov formula.

Ref. \cite{CM} also contains explicit results for the unrefined series for $N_f=N_c+i$ with $i=1,2,3$. Below we will work out the refined
series for the two former cases; our results are consistent with those in \cite{CM} upon refining.

In our case, when $N_f>N_c$ we have an extra constraint $\ell(\la) \le N_c$ which prevents us from doing the sum in a straightforward
manner. However, small $N_f-N_c$ can be worked out. Let us start with $N_f=N_c+1$. We can write the refined series as a sum of two contributions:
\begin{eqnarray} \label{eq:Ncplus1}
  g_{N_c+1,U(N_c)}(t,\tilde t,x,y)
&=&\sum_{\la:\ell(\la) \le N_c} (t\tilde t)^{|\la|} s_\la(\tilde x)  s_\la(\tilde y) \nonumber \\
&=&\sum_{\la} (t\tilde t)^{|\la|} s_\la(\tilde x)  s_\la(\tilde y) -\sum_{\la:\ell(\la) = N_c+1} (t\tilde t)^{|\la|} s_\la(\tilde x)  s_\la(\tilde y) \ .
\end{eqnarray}
The first term can be readily summed, and with the second one we proceed as follows. Since $\ell(\la)=N_f=N_c+1$, all column heights $\la_i \ge 1$, with $i=1,\ldots,N_f$. Therefore there exists a partition $\kappa$ such that $\kappa+1=\la$, which by definition means that $\kappa_i+1=\la_i$ for all $i=1,\ldots,N_f$. For all partitions $\kappa$, the shifted partition $\kappa+1$ is included in the sum of the second term since $\ell(\kappa+1)=N_f$. We also notice that  $s_{\kappa+1}(\tilde x) = (\prod_i \tilde  x_i) s_\kappa(\tilde x) = s_\kappa(\tilde x) $
by Eqs.~\eqref{eq:xprodid} and~\eqref{eq:shiftid}. Collecting these observations, we get
\begin{eqnarray} \label{eq:Ncplus1int}
\sum_{\la:\ell(\la) = N_c+1} (t\tilde t)^{|\la|} s_\la(\tilde x)  s_\la(\tilde y) &=& \sum_\kappa (t\tilde t)^{|\kappa+1|} s_{\kappa+1}(\tilde x)  s_{\kappa+1}(\tilde y) \nonumber \\
=(t\tilde t)^{N_c+1}  \sum_\kappa (t\tilde t)^{|\kappa|} s_{\kappa}(\tilde x)  s_{\kappa}(\tilde y)  &=&  (t\tilde t)^{N_c+1}  \prod_{i,j=1}^{N_c+1}\frac{1}{1-t\tilde t \tilde x_i \tilde y_j} \ .
\end{eqnarray}
Therefore we find
\begin{equation}
 \label{eq:Ncplus1res}
  g_{N_c+1,U(N_c)}(t,\tilde t,x,y)
= \left(1-(t\tilde t)^{N_c+1} \right)\prod_{i,j=1}^{N_c+1}  \frac{1}{1-t\tilde t \tilde x_i \tilde y_j} \ .
\end{equation}

The case $N_f=N_c+2$ can also be treated in a similar manner, but some extra tricks are necessary. The calculation is done in Appendix~\ref{Apprefres}.
We find
\begin{eqnarray} \label{eq:Nfp2res}
&&   g_{N_c+2,U(N_c)}(t,\tilde t,x,y) =  \prod_{i,j=1}^{N_f}\frac{1}{1-t\tilde t \tilde x_i \tilde y_j}\\ \nn
&&\times\Bigg[1-(t\tilde t)^{N_f} -(t\tilde t)^{N_c+1} \sum_{k,\ell=1}^{N_f} \frac{\prod_{i=1}^{N_f}(1-t\tilde t \tilde x_k \tilde y_i)\prod_{j=1}^{N_f}(1-t\tilde t \tilde x_j \tilde y_\ell)}{\tilde x_k^2 \tilde y_\ell^2 (1-t\tilde t \tilde x_k \tilde y_\ell)\prod_{i\ne k} (\tilde x_i-\tilde x_k) \prod_{j\ne \ell} (\tilde y_j-\tilde y_\ell)} \Bigg] \ .
\end{eqnarray}
It may be possible to simplify the sum in the square brackets further. One can also derive similar results for higher fixed $N_f-N_c$, but the expressions quickly become very cumbersome.

\subsection{Results for the refined $SU(N_c)$ series}\label{sec:resultsforSUNc}

The above analysis can be repeated to a large extent in the case of $SU(N_c)$ SQCD.
Now the integrand in Eq.~\eqref{eq:SULagrexp} contains
\begin{equation}
 z_1^k\cdots z_{N_c}^k\prod_{i=1}^{N_f}\prod_{a=1}^{N_c}\frac{1}{\left(1-t \tilde x_i z_a^{-1}\right)\left(1-\tilde t \tilde y_i z_a\right)}= \sum_{\la,\kappa} z_1^k\cdots z_{N_c}^k t^{|\la|} s_\la(\tilde x) s_\la(\bar z)\tilde  t^{|\kappa|} s_\kappa(\tilde y) s_\kappa(z) \ .
\end{equation}
We proceed by using the identity~\eqref{eq:shiftid}. For $k\ge 0$, we have
\beq
 z_1^kz_2^k\cdots z_{N_c}^k s_\kappa(z) = s_{\kappa+k}(z)  \qquad (k\ge 0) \ ,
\eeq
where $\kappa+k$ is defined by $(\kappa+k)_i=\kappa_i+k$  with $i=1,\ldots,N_c$. For $k<0$ we use instead
\beq
 z_1^kz_2^k\cdots z_{N_c}^k s_\la(\bar z) = s_{\la+|k|}(\bar z) \qquad (k< 0) \ .
\eeq
By combining the above equations, and by using orthogonality we find
\beqa \label{eq:Ik}
  &&g_{N_f,SU(N_c)}(t,\tilde t,x,y) = \sum_{k=-\infty}^\infty I_k \\
 I_k&=&\frac{1}{N_c!}\prod_{a=1}^{N_c}  \int_0^{2\pi}\frac{d\tau_a}{2\pi}|\Delta(z)|^2 z_1^kz_2^k\cdots z_{N_c}^k\prod_{i=1}^{N_f}\prod_{a=1}^{N_c}\frac{1}{\left(1-t \tilde x_i z_a^{-1}\right)\left(1-t \tilde y_i z_a\right)} \\ \nn
 &=& \left\{\begin{array}{rcl}
     \displaystyle    \sum_{\la,\kappa}  t^{|\la|} s_\la(\tilde x) \tilde  t^{|\kappa|} s_\kappa(\tilde y) \delta_{\la,\kappa+k(N_c)} \ ; \mbox{\qquad} & k\ge 0 \ , \\
   \displaystyle      \sum_{\la,\kappa}  t^{|\la|} s_\la(\tilde x) \tilde  t^{|\kappa|} s_\kappa(\tilde y) \delta_{\la+|k|(N_c),\kappa} \ ; \mbox{\qquad} & k < 0 \ .
     \end{array}\right.
\eeqa
Recall that the partitions must satisfy $\ell(\la),\ell(\kappa)\le\min(N_f,N_c)$. We can now analyze separately different cases:
\begin{itemize}

 \item $N_f<N_c$: In this case the Kronecker delta symbols in~\eqref{eq:Ik} can only be nonzero for $k=0$.\footnote{The partitions are constrained by $\ell(\la),\ell(\kappa) \le N_f<N_c$, but $\ell(\kappa+k(N_c))=N_c$ for $k>0$ and  $\ell(\la+|k|(N_c))=N_c$ for $k<0$. Therefore, the Kronecker deltas vanish, unless $k=0$.}  Consequently, the integrals $I_k$ vanish for $k \ne 0$, and the result for $g$ is the same as for $U(N_c)$ SQCD,
\beq\label{eq:SUNcNflessthanNc}
 g_{N_f,SU(N_c)}(t,\tilde t,x,y) = I_0 =  \prod_{i=1}^{N_f} \prod_{j=1}^{N_f}\frac{1}{1-t\tilde t \tilde x_i \tilde y_j} \ \qquad (N_f < N_c)
 \ .
\eeq

 \item $N_f=N_c$: Now the Kronecker delta symbols in \eq{Ik} can be nonzero for all values of $k$. We get
\beqa
I_k &=& \sum_\kappa t^{|\kappa|+N_c k} s_{\kappa+k}(\tilde x) \tilde  t^{|\kappa|} s_\kappa(\tilde y) =t^{N_c k}\tilde x_1^k \cdots \tilde x_{N_c}^k \sum_\kappa t^{|\kappa|} s_{\kappa}(\tilde x) \tilde  t^{|\kappa|} s_\kappa(\tilde y) \nonumber\\
 &=& t^{N_c k}  \prod_{i=1}^{N_c} \prod_{j=1}^{N_c}\frac{1}{1-t\tilde t \tilde x_i \tilde y_j} \ ; \qquad k\ge 0 \\
I_k &=&  \sum_{\la}  t^{|\la|} s_\la(\tilde x) \tilde  t^{|\la|+N_c|k|} s_{\la+|k|}(\tilde y) =  \tilde  t^{N_c|k|} \tilde y_1^{|k|} \cdots \tilde y_{N_c}^{|k|} \sum_{\la}  t^{|\la|} s_\la(\tilde x)t^{|\la|} s_{\la+|k|}(\tilde y) \nonumber\\
&=& \tilde  t^{N_c|k|}  \prod_{i=1}^{N_c} \prod_{j=1}^{N_c}\frac{1}{1-t\tilde t \tilde x_i \tilde y_j} \ ; \qquad k < 0 \ ,
\eeqa
where we recalled the definitions of $\tilde x$ and $\tilde y$ in terms of $x$ and $y$. The sum over $k$ can also be done explicitly, giving
\beqa
 & & g_{N_c,SU(N_c)}(t,\tilde t,x,y) \nonumber\\
 & = & \sum_{k=-\infty}^\infty I_k = \left[\frac{1}{1- t^{N_c }} + \frac{\tilde  t^{N_c}}{1-\tilde  t^{N_c}} \right] \prod_{i=1}^{N_c}\prod_{j=1}^{N_c} \frac{1}{1-t\tilde t \tilde x_i \tilde y_j} \nonumber\\
 & = & \frac{1-(t\tilde t)^{N_c}}{\left(1- t^{N_c }\right)\left(1-\tilde  t^{N_c}\right)} \prod_{i=1}^{N_c}\prod_{j=1}^{N_c} \frac{1}{1-t\tilde t \tilde x_i \tilde y_j}  
 \ .\label{eq:SUNcNfequalsNc}
\eeqa
Setting here the $x$'s and $y$'s to unity we recover the unrefined series calculated in \cite{CM}.

\item $N_f=N_c+1$: This special case is worked out in Appendix~\ref{Apprefres}. We find
\begin{eqnarray} \label{eq:SUNcp1res}
&& g_{N_c+1,SU(N_c)}(t,\tilde t,x,y) =  \prod_{i,j=1}^{N_f} \frac{1}{1-t\tilde t \tilde x_i \tilde y_j} \\ \nonumber
&& \times\! \sum_{k,\ell=1}^{N_f}\! \frac{(1-(t\tilde t)^{N_c}/(\tilde x_k\tilde y_\ell))\prod_{i=1}^{N_f}(1-t\tilde t \tilde x_k \tilde y_i)\prod_{j=1}^{N_f}(1-t\tilde t \tilde x_j \tilde y_\ell)}{\tilde x_k \tilde y_\ell(1-t^{N_c}/\tilde x_k)(1-\tilde t^{N_c}/\tilde y_\ell) (1-t\tilde t \tilde x_k \tilde y_\ell)\prod_{i\ne k} (\tilde x_i-\tilde x_k) \prod_{j\ne \ell} (\tilde y_j-\tilde y_\ell)}
\end{eqnarray}

\item $N_f>N_c$ with generic $N_f$: In this case all partitions in \eq{Ik} have lengths less than equal to $N_c$. We can do the sums over the partitions by using the Kronecker deltas:
\beqa
 I_k \label{eq:IkNfgNc}
 &=& \left\{\begin{array}{rcl}
     \displaystyle   t^{kN_c}\!\!\sum_{\kappa:\ell(\kappa)\le N_c}\!\!  t^{|\kappa|} s_{\kappa+k(N_c)}(\tilde x) \tilde  t^{|\kappa|} s_\kappa(\tilde y)  \ ; \mbox{\qquad} & k\ge 0 \ , \\
      \displaystyle  \tilde t^{|k|N_c}\!\!\sum_{\la:\ell(\la)\le N_c}\!\!  t^{|\la|} s_\la(\tilde x) \tilde  t^{|\la|} s_{\la+|k|(N_c)}(\tilde y)  \ ; \mbox{\qquad} & k < 0 \ .
     \end{array}\right.
\eeqa
Proceeding further is difficult, and we will later find an 
interpretation by using the log-gas approach, with qualitatively different
behavior separated by the $N_f=N_c$ threshold. 

\end{itemize}

\subsection{The asymptotics of the unrefined Hilbert series}

Let us then discuss the asymptotics of the unrefined Hilbert series and the singularities of their sums for $U(N_c)$ and $SU(N_c)$ theories.
As the series for $N_f \le N_c$ were solved explicitly above, their singularities are readily known, and we can restrict to $N_f \ge N_c$. For the unrefined series,
the results (\ref{eq:Ungen}),~\eqref{eq:Ik}, and (\ref{eq:IkNfgNc}) may be written as
\begin{eqnarray} \label{eq:unrefdef}
 g_{N_f,U(N_c)}(t,\tilde t) &=& \sum_{\la:\ell(\la) \le N_c} (t\tilde t)^{|\la|}d_\la^2\\ \nonumber
 g_{N_f,SU(N_c)}(t,\tilde t) &=& \sum_{k=0}^\infty t^{kN_c} \sum_{\kappa:\ell(\kappa) \le N_c} (t\tilde t)^{|\kappa|}d_{\kappa+k(N_c)} d_\kappa \\ \nn
&& \ \ + \sum_{k=-\infty}^{-1} \tilde t^{|k|N_c}  \sum_{\la:\ell(\la) \le N_c} (t\tilde t)^{|\la|}d_\la d_{\la+|k|(N_c)} \ ,
\end{eqnarray}
where $d_\la =s_\la(1,1,\ldots,1)$ denotes the dimension of the $SU(N_f)$ representation characterized by $\la$.

The asymptotics of the series and correspondingly the singularities of the series can be found by a ``brute force'' calculation,
which is sketched in Appendix~\ref{Appdasympt}. We find that for large $k$ and $s=|\la|$,
\begin{equation}
 G(s,k) \equiv \sum_{\la: |\la|=s,\ \ell(\la) \le N_c} d_{\la+k(N_c)} d_\la \sim (s+\# k)^{N_c(N_f-N_c)} s^{N_fN_c-1} \ .
\end{equation}
Consequently, we get
\begin{equation} \label{eq:UNasympt}
 g_{N_f,U(N_c)}(t,\tilde t) = \sum_{s=0}^\infty(t \tilde t)^s G(s,0) \sim \sum_{s=0}^\infty (t \tilde t)^s s^{2N_fN_c-N_c^2-1} \sim \frac{1}{(1-t\tilde t)^{2N_fN_c-N_c^2}}
\end{equation}
for $N_f \geq N_c$ as $t\tilde t \to 1$, where the order of the singularity matches with the expected dimension of the moduli space.

In the $SU(N_c)$ case, let us first analyze the two terms of Eq.~\eqref{eq:unrefdef} separately.
That is, we first write
\begin{equation} \label{eq:gSUdec}
 g_{N_f,SU(N_c)}(t,\tilde t) = \sum_{k=0}^\infty \sum_{s=0}^\infty t^{k N_c} (t\tilde t)^s G(s,k) + \sum_{k=-\infty}^{-1} \sum_{s=0}^\infty \tilde t^{|k| N_c} (t\tilde t)^s   G(s,|k|) \ .
\end{equation}
Here the first term is singular at $t \to 1$ and at $t \tilde t \to 1$. In the first case we find
\begin{eqnarray} \label{eq:SUsumdivs}
  \sum_{k=0}^\infty \sum_{s=0}^\infty t^{k N_c} (t\tilde t)^s G(s,k) &\mathop{\sim}\limits_{t \to 1}&  \frac{1}{(1-t)^{N_c(N_f-N_c)+1}} \sum_s (t \tilde t)^s s^{N_cN_f-1} \nonumber \\
&\mathop{\sim}\limits_{t \tilde t \to 1}&  \frac{1}{(1-t)^{N_c(N_f-N_c)+1}} \frac{1}{(1-t\tilde t)^{N_cN_f}} 
\end{eqnarray}
where the first line gives the order of the pole at $t=1$, and the second line adds the leading behavior of its residue as we further take $t \tilde t \to 1$. For the other singularity we find similarly
\begin{eqnarray} \label{eq:SUsumdivs2}
 \sum_{k=0}^\infty \sum_{s=0}^\infty t^{k N_c} (t\tilde t)^s G(s,k) &\mathop{\sim}\limits_{t \tilde t \to 1}& \frac{1}{(1-t\tilde t)^{2 N_c N_f-N_c^2}} \sum_k t^{k N_c} \nonumber\\
&\mathop{\sim}\limits_{t \to 1}& \frac{1}{(1-t\tilde t)^{2 N_c N_f-N_c^2}} \frac{1}{1-t} \ ,
\end{eqnarray}
so that the result depends on the order of limits. For the second term in Eq.~\eqref{eq:unrefdef} we find analogous results with $t \leftrightarrow \tilde t$.

Adding the results up, we find
\begin{eqnarray} \label{eq:SUt1sing}
 g_{N_f,SU(N_c)}(t,\tilde t) &\mathop{\sim}\limits_{t \to 1}&  \frac{1}{(1-t)^{N_c(N_f-N_c)+1}} \nn\\
 g_{N_f,SU(N_c)}(t,\tilde t) &\mathop{\sim}\limits_{\tilde t \to 1}&  \frac{1}{(1-\tilde t)^{N_c(N_f-N_c)+1}}
\end{eqnarray}
where $N_f \geq N_c$.\footnote{Here we only wrote down the order of the pole and left out the structure of the residue, which is given in Eq.~\eqref{eq:SUsumdivs} as $t \tilde t \to 1$.}  We are also tempted to write down the result as $t \tilde t \to 1$. However, the corresponding singularity appears at the same order in both terms of Eq.~\eqref{eq:gSUdec}, and since taking $t\tilde t \to 1$ fixes $t = 1/\tilde t$, cancellations between the two terms can, and typically do, take place. Therefore, we cannot predict the order of the singularity at $t \tilde t \to 1$, expect that it is smaller or equal to $2N_cN_f-N_c^2$.

Finally, let us calculate the order of the ``overall'' singularity, taking first $\tilde t =t$ and then $t \to 1$. In this case, the two terms in Eq.~\eqref{eq:gSUdec} are identical for $|k|\ge 1$ and add up without the possibility of any cancellations. As seen from Eqs.~\eqref{eq:SUsumdivs} and~\eqref{eq:SUsumdivs2} by setting $\tilde t =t$, or by direct calculation,
\begin{eqnarray} \label{eq:SUtt1sing}
 g_{N_f,SU(N_c)}(t,\tilde t=t) &\mathop{\sim}\limits_{t \to 1}& \frac{1}{(1-t)^{2N_fN_c-N_c^2+1}}
\end{eqnarray}
so that the order of the pole is indeed the dimension of the SQCD moduli space. The extra factor of $(1-t)$ with respect to the $U(N_c)$ result arises from the additional sum over $k$.

\section{Hilbert series in the Veneziano limit}

There are two motivations to consider the Veneziano limit $N_c,N_f\gg 1$ with fixed ratio $N_f/N_c$. On one hand
we can expect to find simplifications for $N_f>N_c$ where only a few explicit results are known.
This was also done
in \cite{CM} and it is interesting to compare the results.
On the other hand, we will use a different method where we reinterpret the Hilbert series as a canonical ensemble
partition function of a two-dimensional gas of point particles confined on a unit circle with an additional
external charge. The system is a specific case of a Coulomb gas; we will
follow a random matrix theory literature convention and refer to
it as a log-gas (due to the logarithmic Coulomb potential in two dimensions), see \cite{mehta,Forresterbook}.
This method will bring new physical insight to the threshold $N_f=N_c$.

\subsection{The log-gas approach and the $U(N_c)$ Hilbert series}

We begin by rewriting the $U(N_c)$ unrefined Hilbert series as a log-gas partition function. 
From the series expansion \eq{Ungen} we observe that $g_{N_f,U(N_c)}(t,\tilde t,x,y)$ and thus also the unrefined series  $g_{N_f,U(N_c)}(t,\tilde t)$ depend on $t$ and $\tilde t$ only through the combination $t\tilde t$. Therefore, we can take $t=\tilde t$ without loss of generality. The defining integral can be written as
\beqa
 g_{N_f,U(N_c)}(t) &=& \frac{1}{N_c!}\prod_{a=1}^{N_c}  \int_0^{2\pi}\frac{d\tau_a}{2\pi}|\Delta(z)|^2 \prod_{a=1}^{N_c}\frac{1}{\left(z_a-t \right)^{N_f}\left(\bar z_a-t \right)^{N_f}} \\
&=& \frac{1}{N_c!}\prod_{a=1}^{N_c}  \int_0^{2\pi}\frac{d\tau_a}{2\pi}|\Delta(z)|^2 \prod_{a=1}^{N_c}\left|z_a-t \right|^{-2N_f} \ .
\eeqa
We restrict to the region $0 \le t<1$ where this integral and the Hilbert series converge.
We can rewrite the integral in the form of a canonical partition function
\be \label{eq:Zredef}
g_{N_f,U(N_c)}(t) = \frac{1}{N_c!} \prod_{a=1}^{N_c}\int_0^{2\pi}\frac{d\tau_a}{2\pi} e^{-2 H}\ ,
\ee
where the Hamiltonian $H$ is
\be \label{eq:Ham}
 H = - \sum_{1\le a<b\le N_c} \log |e^{i\tau_a}-e^{i\tau_b}| + N_f\sum_{a=1}^{N_c} \log |e^{i\tau_a}-t|  \ ,
\ee
and we fixed the inverse temperature to $\beta=2$. The Hamiltonian describes an ensemble of $N_c$ point
particles (of charge +1) on the circle at $e^{i\tau_1},\ldots,e^{i\tau_{N_c}}$,
which interact via the Coulomb potential, with an external particle with charge $-N_f$, located on the real line,
inside the unit disk at $t\in (0,1)$ \cite{VEPA,Balasubramanian:2006sg}. The system is symmetric with respect to an overall sign flip of the charges.

We shall now apply the general method described in \cite{vikareview,PNG}; the
case with one external charge inside the disk was studied in detail in \cite{bulk2pt}. The same method was used to analyze unitary matrix models with different potentials in \cite{Mandal}. We
repeat the main points.
The leading contribution to $\log g_{N_f,U(N_c)}(t)$ in the Veneziano limit of large $N_{c,f}$ with fixed $N_f/N_c\equiv r$ may be
calculated by going to the continuum limit of log-gas. Calculating the partition function in this limit
resolves into solving a potential problem with the external charge at the point $z=t$. There is an interesting transition point \cite{bulk2pt}: because all
particles on the circle carry the same charge, the external charge may cause a gap in the charge distribution on the circle if it comes sufficiently close to it.
Therefore, we actually need to solve two potential problems: one with a gap and one without.
In comparing the analysis with that in \cite{bulk2pt}, notice that
because the sign of the charge of the external particle is now opposite to that of the distribution on
the circle, their
Coulomb interaction is attractive, and the gap will appear at the opposite side.

As pointed out above, there will be a threshold for the gap formation at $N_f=N_c$, which is qualitatively understood as follows. As the external charge approaches the boundary, it is able to attract at most $N_f$ positive unit charges in its vicinity, because these charges will screen the external charge so that its potential effectively vanishes at long distances. In particular, if $N_f>N_c$, all unit charges are attracted by the external charge and a gap eventually forms opposite to it. For $N_c>N_f$, the leftover $N_c-N_f$ charges will remain evenly distributed  on the unit circle, and no gap is formed. This behavior will be confirmed by the explicit calculation below.

\paragraph{No gap in the distribution.}
The solution for a potential on the disk with one charge is well-known,
\beq
 V(z) = N_f\log\frac{z-t}{1-zt} \ .
\eeq
The charge distribution on the boundary of the disk (with $z=e^{i\tau}$) is given, up to a constant, by the radial electric field \cite{bulk2pt}:
\beq
 \rho(\tau) = \mathrm{const} + \frac{1}{2\pi}|V'(z)|_{z=e^{i\tau}} = \frac{1}{2\pi}\left(N_c-N_f +N_f\frac{1-t^2}{1+t^2-2 t \cos \tau}\right) \ .
\eeq
Here the constant was fixed by the normalization $\oint d\tau \rho(\tau) = N_c$. The distribution is minimized opposite to the external particle,
\beq
 \min \rho(\tau) = \rho(\pi) = \frac{N_c}{2\pi}\left(1-\frac{2tr}{1+t}\right) \ ,
\eeq
where $r=N_f/N_c$. A gap forms when the minimum value reaches zero. The gapless phase requires
\beq \label{eq:tcdef}
 t < \frac{1}{2r-1} \equiv t_c \ .
\eeq
Note that for $N_f <N_c$, $|t_c|>1$ so that the system is always in the gapless phase, as expected.
The continuum energy of the system is as in \cite{bulk2pt}
\bea
 \mathcal{E} &=& -\frac{1}{2} \int d\tau_1 d\tau_2\ \rho(\tau_1) \rho(\tau_2) \log\left|e^{i\tau_1}-e^{i\tau_2}\right| + N_f  \int d\tau \ \rho(\tau) \log\left|e^{i\tau}-t\right| \nn\\
 &=& \frac{N_f^2}{2}\log(1-t^2) \label{eq:Enogap} \ .
\eea

The leading result for the unrefined Hilbert series in the gapless phase reads
\beq
\label{eq:lead1}
 g_{N_f,U(N_c)}(t) \simeq e^{-2\mathcal{E}} = \frac{1}{(1-t^2)^{N_f^2}} \ ; \quad \qquad 0 \leq t \leq t_c \ .
\eeq
The constraint for $t$ can be dropped if $N_f<N_c$, and the gapless result holds for all $0 \leq t < 1$. We immediately notice that the order of the pole at $t=1$ agrees with the dimension of the moduli space of $U(N_c)$ SQCD for $N_f<N_c$, and actually the results matches with the exact result in this region (see Eq.~\eqref{eq:unrefUN} and \cite{CM}). 
We shall also compare this to the Schur polynomial result and the results in \cite{CM},
but first we discuss what happens when a gap forms.

\paragraph{One gap in the distribution.}
For $t>t_c$ the charge distribution develops a gap opposite to the external charge. For this to be possible,
we need $N_f>N_c$. The calculation of the electrostatic energy $\mathcal{E}$ is quite technical in the presence of the gap. We only report the results here, as a similar calculation (for opposite bulk charge) was carried out in detail in \cite{bulk2pt}. See also \cite{PNG} for a generic discussion of the solution for arbitrary numbers of gaps.

Following \cite{bulk2pt}, we see that the support of the charge distribution is $[-\tau_c,\tau_c]$. The edges of the gap,  $\pm\tau_c$, are given by
\beq
 \sin^2\frac{\tau_c}{2}= \frac{\frac{r^2}{(r-1)^2}-1}{\frac{(1+t)^2}{(1-t)^2}-1} = \frac{\chi^2-1}{\delta(t)^2-1} \ ,
\eeq
where
\beq
 \chi = \frac{r}{r-1} \ ; \qquad \delta(t) = \frac{1+t}{1-t} \ .
\eeq
Notice that $\tau_c$ is only defined in the one-gap phase where $t>t_c$, or equivalently $\delta(t)>\chi$.
In terms of the new variables, we calculate the potential $U_0$ at the circle\footnote{The potential
is constant, otherwise the charge distribution would not be static.}
\beq
 \frac{U_0}{N_c} = -\frac{1}{2}\log \frac{\chi-1}{\delta(t)-1} - \frac{2r-1}{2}\log \frac{\chi+1}{\delta(t)+1}
\eeq
and the interaction energy between the external particle and the charge distribution
 \beq
\frac{\mE_{c-\xi}}{N_c^2}
= r(1-r)\left[\log\frac{\delta(t)+1}{\chi+1}+\chi\log\frac{4\chi}{(\chi+1)(\delta(t)+1)}\right] \ .
\eeq
Notice, that after the substitution $N_f\to-\xi$ these results are formally identical to the one-gap results of \cite{bulk2pt} where the bulk
external particle had an opposite charge. However, there is an important difference in the definition of $\delta(t)$: in \cite{bulk2pt} $\delta$ went
to zero as the particle was approaching the unit circle, whereas in the present case $\delta$ goes to infinity in this limit.

The total energy becomes
\beqa
 \frac{\mathcal{E}}{N_c^2} &=& -\frac{(2r-1)^2}{4}\log\frac{\chi+1}{\delta(t)+1}- \frac{1}{4}\log\frac{\chi-1}{\delta(t)-1}+\frac{r^2}{2}\log\frac{4 \chi}{(1+\delta(t))^2} \\
&=& \frac{r^2}{2}\log\left(1-t^2\right)-\frac{(2r-1)^2}{4}\log\frac{\chi+1}{\delta(t)+1}- \frac{1}{4}\log\frac{\chi-1}{\delta(t)-1}+\frac{r^2}{2}\log\frac{\chi}{\delta(t)} \label{eq:Egap}
\eeqa
giving the leading $\mathcal{O}(N_c^2)$ term of $\log g_{N_f,U(N_c)}(t)$ in the gapped phase:
\beqa \label{eq:leadgap}
  g_{N_f,U(N_c)}(t) &\simeq& \left(\frac{\chi+1}{\delta(t)+1}\right)^{(2N_f-N_c)^2/2}\left(\frac{\chi-1}{\delta(t)-1}\right)^{N_c^2/2} \nonumber\\
&& \times\left(\frac{\delta(t)}{\chi}\right)^{N_f^2} \frac{1}{(1-t^2)^{N_f^2}}\ \ ; \qquad\quad  t_c\leq t<1  \ .
\eeqa
Notice that the results~\eqref{eq:lead1} and~\eqref{eq:leadgap} join smoothly for $t \to t_c$, as $\delta(t) \to \chi$.
In \cite{PNG} we proved that $\mathcal{O}(N_c)$ corrections to $\log g$ vanish, but there might be nontrivial contributions at $\mathcal{O}(N_c^0)$. Recall also that the result can be extended for unequal $t$ and $\tilde t$ by using the identity $g_{N_f,U(N_c)}(t,\tilde t) = g_{N_f,U(N_c)}(\sqrt{t\tilde t})$, {\em i.e.}, by replacing $t \to \sqrt{t \tilde t}$ in the formulas above.

As $t \to 1$, we get
\beq
\label{eq:lead2}
 g_{N_f,U(N_c)}(t) \sim \left(\frac{\chi+1}{2}\right)^{(2N_f-N_c)^2/2}\left(\frac{\chi-1}{2}\right)^{N_c^2/2} \left(\frac{1}{\chi}\right)^{N_f^2} \frac{1}{(1-t)^{2N_fN_c-N_c^2}} \ .
\eeq
We observe that the exponent in the divergent factor (in the $t\to 1$ limit) is different
from that of (\ref{eq:lead1}), $N^2_f$ has changed to $2 N_fN_c-N_c^2$. This is good,
because the power in the divergent factor then manifestly agrees with the dimension of the moduli space of $U(N_c)$ SQCD, as expected (up to $\mathcal{O}((N_c)^0)$ corrections which vanish in this case).

It is interesting to mention that the phase transition for gap formation is of third order, similarly as in \cite{grosswitten}. This can be easily checked by calculating the
derivatives of $\log g_{N_f,U(N_c)}(t)$ in (\ref{eq:Enogap}) and (\ref{eq:Egap}) with respect to $t$ and noting that the third derivative is discontinuous at $t_c$.

\subsection{Applying the log-gas approach in the $SU(N_c)$ case} 

We argue that the results of the previous sections hold for the $SU(N_c)$ gauge group as well, if $t=\tilde t$, and up to subleading corrections in the Veneziano limit. This is expected, because the definitions of the Hilbert series differ only by the extra condition for the total phase $\sum_a\tau_a=2\pi n$, which appears in the $SU(N_c)$ case. When $N_c$ is large, small modifications of the configuration with $\delta \tau_a \sim 1/N_c$ are enough to make the total phase 
arbitrary while changing the contribution to the partition function by a subleading amount.\footnote{Actually, as we have verified numerically, this naive argument fails for $t \ne \tilde t$ when $t$ or $\tilde t$ is close to one. Notice that we discuss here only the case where $t=\tilde t$ where the log-gas approach is directly applicable. Then, as we shall see below, the $U(N_c)$ and $SU(N_c)$ Hilbert series indeed match in the Veneziano limit.}

An analytic argument can be constructed as follows. We first separate the dependence of the total phase of the $U(N_c)$ partition function:
\bea
  g_{N_f,U(N_c)}(t,\tilde t) &=& \int_{-\pi}^\pi \frac{d\theta}{2\pi} \times \frac{1}{N_c!}\prod_{a=1}^{N_c}  \int_0^{2\pi}\frac{d\tau_a}{2\pi}\sum_{k=-\infty}^\infty 2\pi\delta\left(\sum_a\tau_a-\theta-2\pi k\right) \\\nn
&&\times |\Delta(z)|^2 \prod_{a=1}^{N_c}\frac{1}{\left(z_a-t \right)^{N_f}\left(\bar z_a-\tilde t \right)^{N_f}} \ .
\eea
The overall phase $\theta$ in the delta function can be eliminated by a small overall shift $\tau_a \to \tau_a + \theta/N_c$ of the configuration. By rotational symmetry on the $z$-plane, this can be transferred to small displacements of the ``external particles'' at $z=t$ and $z=\tilde t$, which leads to the relation
\be \label{eq:USUrel}
  g_{N_f,U(N_c)}(t,\tilde t) = \int_{-\pi}^\pi \frac{d\theta}{2\pi} \   g_{N_f,SU(N_c)}(t e^{-i\theta/N_c},\tilde t e^{i\theta/N_c})
\ee
between the partition functions of the two gauge groups.

Notice, however, that in the log-gas calculation above we set $t=\tilde t$. While this could be done for $U(N_c)$ without loss of generality, it is a true extra constraint in the $SU(N_c)$ case. 
For real $t=\tilde t$ the effective Hamiltonian, given by the logarithm of the integrand on the right hand side of Eq.~\eqref{eq:USUrel}, becomes
\bea \label{eq:Ham3}
 H_\mathrm{eff}
 &=&  - \sum_{1\le a<b\le N_c} \log |z_a-z_b| + N_f\sum_{a=1}^{N_c} \log |z_a-t |  + \frac{N_f t \theta}{N_c} \sum_{a=1}^{N_c}\frac{\sin \tau_a}{|z_a-t|^2} + \cdots
\eea
Taking into account the number of terms in the sums, the $\theta$-independent term is $\mathcal{O}(N_c^2)$ which makes it the leading
 one in the Veneziano limit, the linear term in $\theta$ is $\mathcal{O}(N_c)$, and we dropped $\mathcal{O}(N_c^0)$ contributions.

We shall now argue that the next-to-leading $\mathcal{O}(N_c)$ correction vanishes in this case. This seems plausible, since due to symmetry with
respect to the reflections along the real axis, $g_{N_f,SU(N_c)}(t e^{i\theta/N_c})$ must be an even function of $\theta$.
More precise argument is as follows. In \cite{PNG} we showed carefully that (minus two times) the minimal energy of the saddle point
configurations gives the logarithm of the partition function up to $\mathcal{O}(N_c^0)$ terms. Therefore, it is enough to show that
the first variation of the saddle point energy due to the next-to-leading term $\propto \theta$ in Eq.~\eqref{eq:Ham3} vanishes.

First, we note that the variation of the sum of energies of the leading $\mathcal{O}(N_c^2)$ terms, which follows from small changes of $\tau_a$ in the
saddle point configuration due to the additional term, is zero. This is the case because the saddle point energy is unchanged under any variation of $\tau_a$ by definition.
Second, the leading saddle point configuration must be symmetric with respect to the real $z$-axis. Therefore, the
possible $\mathcal{O}(N_c)$ term obtained by evaluating on the saddle point configuration the next-to-leading term in the
Hamiltonian~\eqref{eq:Ham3}, which is odd under the reflection with respect to real axis, vanishes.

We conclude that for $t=\tilde t$ the right hand side of Eq.~\eqref{eq:USUrel} is independent of $\theta$ up to $\mathcal{O}(N_c^0)$ corrections. Therefore
\be \label{eq:USUres}
 \log g_{N_f,U(N_c)}(t) = \log g_{N_f,SU(N_c)}(t) + \mathcal{O}(N_c^0)
\ee
in the Veneziano limit. Therefore, the log-gas result holds up to $\mathcal{O}(N_c^0)$ corrections
for both $U(N_c)$ and $SU(N_c)$.\footnote{Actually, according to the GCBO result below the
corrections in the gapless phase are exponentially suppressed.} However, we shall see
numerically below that the convergence towards the leading term is considerably faster for $U(N_c)$ than for $SU(N_c)$.

\subsection{The GCBO approach}

Let us now compare our results with those in \cite{CM}, which
studied the behavior of the unrefined Hilbert series in the regime $r=N_f/N_c>1$ 
by using the Geronimo-Case-Borodin-Okounkov (GCBO) formula for Toeplitz determinants \cite{GCBO}.
Their result reads
\beqa \label{eq:GCBO1}
  g_{N_f,U(N_c)}(t,\tilde t) &\simeq& \frac{1-{r N_c\choose N_c+1}^2 \mathcal{F}(N_c,r,t\tilde t)^2 (t\tilde t)^{N_c+1}}{(1-t\tilde t)^{r^2N_c^2}} \\
g_{N_f,SU(N_c)}(t,\tilde t) &\simeq& \frac{1}{(1-t\tilde t)^{r^2N_c^2}} \left[1 +{r N_c\choose N_c} \widehat{\mathcal{F}}(N_c,r,t\tilde t) (t^{N_c}+\tilde t^{N_c})\right. \nonumber\\
&&-{r N_c\choose N_c+1}^2 \mathcal{F}(N_c,r,t\tilde t)^2 (t\tilde t)^{N_c+1}\bigg] \ ,  \label{eq:GCBO2}
\eeqa
where the numerators come from an expansion of the important determinant factor in the GCBO formula, and
\beqa \label{eq:Fdefs}
\mathcal{F}(N_c,r,t\tilde t)&=& _2F_1(-(r-1)N_c+1,rN_c;N_c+2;t\tilde t) \\
  \widehat{\mathcal{F}}(N_c,r,t\tilde t) &=&  _2F_1(-(r-1)N_c+1,rN_c;N_c+1;t\tilde t) \ .
\eeqa
These expressions are expected to be good approximations for large $N_c$ and whenever $t\tilde t (r-1)^2$ is small. If $r$ is fixed and $t \tilde t$ is small, which is the limit we will be using here, $N_c$ does not need to be large.

In the limit of large $N_c$, we expect the GCBO determinant factor, which gave the numerator in Eqs.~\eqref{eq:GCBO1}, 
to contribute to the leading divergent
behavior as $t,\tilde t \to 1$, so that the 
order of the divergence
will correctly match with the dimension of the moduli space. This cannot be however achieved with the above approximation which is restricted for small $t\tilde t$.

Ref. \cite{CM} pointed out the need for
good asymptotic formulas for the hypergeometric function. We study this by applying
the saddle point method to the integral representation of the hypergeometric function $_2F_1$ and
draw some lessons from the analysis. This analysis will show how the result~\eqref{eq:GCBO1} behaves in the Veneziano limit and for small $t \tilde t$, where it is expected to be applicable.

The saddle point analysis is detailed in Appendix~\ref{AppGCBO}. 
We find the following leading order results for the correction terms in Eqs.~\eqref{eq:GCBO1} and~\eqref{eq:GCBO2}:
\begin{eqnarray} \label{eq:spres}
  {r N_c\choose N_c} \widehat{\mathcal{F}}(N_c,r,t\tilde t) (t\tilde t)^{N_c/2} &\sim& {r N_c\choose N_c+1} \mathcal{F}(N_c,r,t\tilde t) (t\tilde t)^{N_c/2}  \\\nonumber
 &\sim& \left\{\begin{array}{rcl}
     \displaystyle  e^{N_c \left[F(s_-)+\inv{2}\log t\tilde t\right]}  \ ; \mbox{\qquad} & t\tilde t<(t\tilde t)_c  \\
   \displaystyle    \cos(N_c B+\phi) \ ; \mbox{\qquad} & t\tilde t > (t\tilde t)_c \ ,
     \end{array}\right.
\end{eqnarray}
where
\begin{eqnarray}
  (t\tilde t)_c &=& 1/(2r-1)^2 \\
  F(s) &=& \left[r\log s-(r-1)\log(s-1)+(r-1)\log(1-sz)\right] \\
\label{eq:spnogap}  s_\pm &=& 1 +\frac{1-t\tilde t\pm\sqrt{\left(1-t\tilde t\right)\left[1-(2r-1)^2t\tilde t \right]}}{2rt\tilde t} \ ; \qquad t\tilde t<(t\tilde t)_c \\
\label{eq:spgap} \hat s_\pm &=&  1 +\frac{1-t\tilde t\pm i \sqrt{\left(1-t\tilde t\right)\left[(2r-1)^2t\tilde t-1 \right]}}{2rt\tilde t} \ ; \qquad t\tilde t>(t\tilde t)_c \\
 B &=& \frac{1}{2i}\left(F(\hat s_+)-F(\hat s_-)\right) = \mathrm{Im} F(\hat s_+) 
\end{eqnarray}
and the order one phase $\phi$ is not determined at leading order (see Appendix~\ref{AppGCBO} for the result at next-to-leading order).
We see that there is a transition in the behavior of the correction terms, which takes place at $t\tilde t = 1/(2r-1)^2$.
As the reader may have already noticed, the critical value matches with that of the log-gas
analysis (see Eq.~\eqref{eq:tcdef}), where a gap appeared in the charge distributions.
In the saddle point analysis, the transition arises as the saddle points in the integral representation of the
hypergeometric functions move from the real line into the complex plane, as can be
seen from Eqs.~\eqref{eq:spnogap} and~\eqref{eq:spgap}, which give the saddle points in the two regions (see Appendix~\ref{AppGCBO} for details).

\begin{figure}[!htb]
\centering
\includegraphics[width=0.49\textwidth]{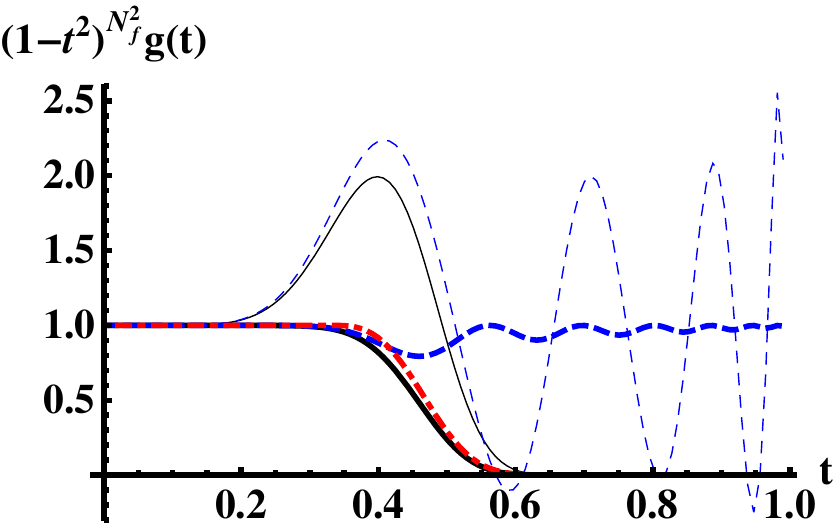}\hspace{2mm}%
\includegraphics[width=0.49\textwidth]{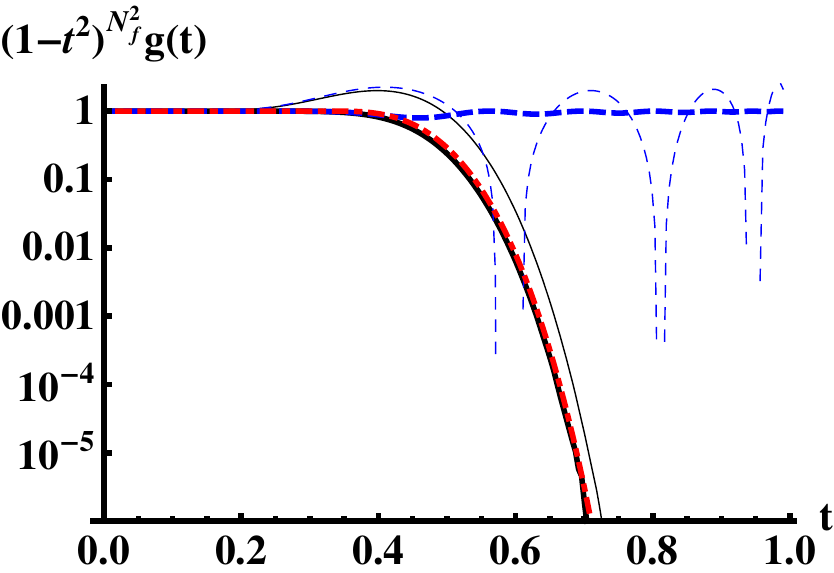}

\vspace{4mm}

\includegraphics[width=0.49\textwidth]{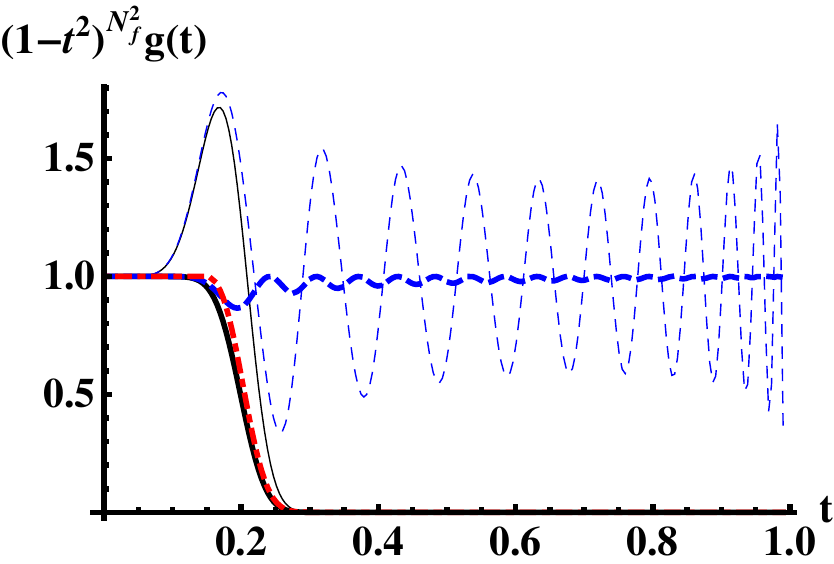}\hspace{2mm}%
\includegraphics[width=0.49\textwidth]{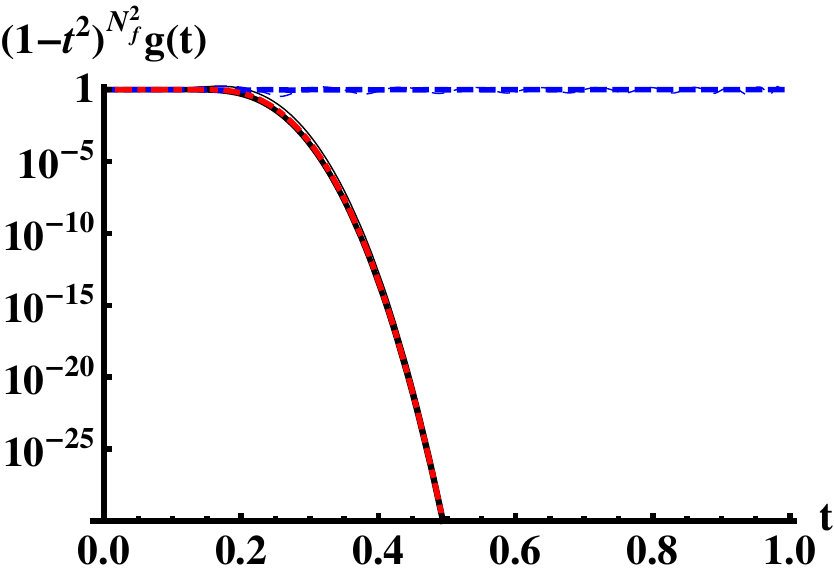}
\caption{\label{figcomp}Comparison of the various approximations and exact results for $(1-t^2)^{N_f^2}g_{N_f,U(N_c)}(t)$ and $(1-t^2)^{N_f^2}g_{N_f,SU(N_c)}(t)$ as a function of $t$. Top row: $N_c=8$ and $N_f=16$. Bottom row: $N_c=8$ and $N_f=32$. Left column: linear scale. Right column: logarithmic scale. Thick (thin) lines are the results for the $U(N_c)$ ($SU(N_c)$) gauge group. Solid black lines are the exact results. Dashed blue lines give the GCBO result. The (single) dotdashed red line is the result in the Veneziano limit (given by the log-gas approach).}
\end{figure}

The saddle point result~\eqref{eq:spres} for the correction terms in the GCBO approach behaves as follows.
For small $t\tilde t$ the factor $F(s_-)+\inv{2}\log t\tilde t$ is negative so that the corrections are highly
suppressed. As $t\tilde t$ approaches the critical value, $F(s_-)+\inv{2}\log t\tilde t$ approaches zero, and
hence the corrections become large. For $t\tilde t$ larger than the critical value, the corrections are $\mathcal{O}(1)$ and
oscillate wildly. This signals the fact that using only the first terms of the GCBO formula, which
gave the results~\eqref{eq:GCBO1} and~\eqref{eq:GCBO2}, is not sufficient to accurately describe the Hilbert series in this region.

Finally, we compare the various results in Fig.~\ref{figcomp} for $t = \tilde t$ and for two choices of $N_f$ and $N_c$.
We also included the exact result, which can be quite easily evaluated numerically by using Toeplitz
determinants of Appendix~\ref{AppToeplitz} up to $N_c \sim 10$.
The results are in line
with our expectations. Generally, the GCBO result is a very good approximation at small $t$, but fails to
reproduce the large $t$ behavior. The leading term in the Veneziano limit, evaluated by mapping to log-gas, does
not work as well for low $t$, but continues to be a reasonable approximation to the exact
result all the way to $t=1$. The GCBO approach seems to work better for $SU(N_c)$ than for $U(N_c)$, while for the log-gas result the opposite is true.

\bigskip
\noindent

{\bf \large Acknowledgments}

We wish to thank, Takeshi Morita, Vasilis Niarchos, and Diego Rodr\'iguez-G\'omez for discussions and Noppadol Mekareeya for correspondence.
N.J. is supported by the MICINN and FEDER (grant FPA2008-01838), the
Spanish Consolider-Ingenio 2010 Programme CPAN (CSD2007-00042), and the Xunta de
Galicia (Conselleria de Educacion and grant INCITE09-206-121-PR). N.J. wishes to thank University of Porto for hospitality while this work was in progress. 
M.J. has been  supported in part by the European grants FP7-REGPOT-2008-1:  CreteHEPCosmo-228644 and PERG07-GA-2010-268246.
E.K-V. has been supported in part by the Academy of Finland grant number~1127482.

\appendix

\section{Hilbert series as Toeplitz determinants} \label{AppToeplitz}

In this Appendix we discuss the relation between the Hilbert series and Toeplitz determinants, pointed out in \cite{CM}. In the definition of the (refined) Hilbert series for the $U(N_c)$ gauge group,
\begin{equation} \label{eq:refdefapp}
 g_{N_f,U(N_c)}(t,\tilde t,x,y) = \frac{1}{N_c!}\prod_{a=1}^{N_c}  \int_0^{2\pi}\frac{d\tau_a}{2\pi}|\Delta(z)|^2 \prod_{i=1}^{N_f}\prod_{a=1}^{N_c}\frac{1}{\left(1-t \tilde x_i z_a^{-1}\right)\left(1- \tilde t \tilde y_i z_a\right)} \ ,
\end{equation}
we may expand
\be
f(z_a)=\prod_{i=1}^{N_f} \frac{1}{\left(1-t \tilde x_i z_a^{-1}\right)\left(1- \tilde t \tilde y_i z_a\right)} = \sum_{n=-\infty}^\infty \hat f_n(t,\tilde t,x,y) e^{i n \tau_a}
\ee
(recall that $z_a= e^{i\tau_a}$). Then an easy computation (see, {\emph{e.g.}}, \cite{CM}) shows that
\be \label{eq:detTres}
 g_{N_f,U(N_c)}(t,\tilde t,x,y) = \det T[f] \equiv \det \left( \hat f_{i-j}\right)_{i,j=1,\ldots,N_c} \ .
\ee
While the Fourier coefficients $\hat f_n$ are quite complicated for the refined Hilbert series, in the unrefined case (setting all $x_i$ and $y_j$ to unity) we find an explicit formula:
\be \label{eq:fnres}
 \hat f_n(t,\tilde t) = (-1)^n {-N_f \choose |n|}\, _2F_1(N_f+|n|,N_f,|n|+1; t\tilde t) \times
\left\{\begin{array}{cc}
 \tilde t^n & \qquad n \geq 0 \\
  t^{-n} &\qquad n<0
\end{array}\right. \ .
\ee
Using this in Eq.~\eqref{eq:detTres} gives explicit results for the unrefined Hilbert series for small $N_c$ (and arbitrary $N_f$) which we shall use to numerically evaluate the Hilbert series exactly in some examples.

Recall from Eq.~\eqref{eq:SULagrexp} that the Hilbert series for $SU(N_c)$ could be written as a sum over integrals $I_k$ otherwise similar to \eqref{eq:refdefapp}, but containing additional phase factors $z_1^k \cdots z_{N_c}^k$. These modify the result~\eqref{eq:detTres} to read
\be \label{eq:Ikdet}
 I_k(t,\tilde t,x,y) = \det \left( \hat f_{i-j-k}\right)_{i,j=1,\ldots,N_c} \ ,
\ee
{\emph{i.e.}}, there is a shift by $k$ in the indices of the Fourier coefficients. Again, the result~\eqref{eq:fnres} may be used to evaluate explicitly the integrals $I_k(t,\tilde t)$, appearing in the expansion of the unrefined Hilbert series, for small $N_c$.

Finally, we notice that the integrals for the unrefined series satisfy
\be
 I_k(t,\tilde t) = t^k \hat I_k(t\tilde t) \ ,
\ee
as can be verified by studying the determinant~\eqref{eq:Ikdet}, or by studying the transformation integrals~\eqref{eq:SULagrexp} under rotations around the origin. In particular, the unrefined $U(N_c)$ Hilbert series (which equals $I_0$) depends on $t$ and $\tilde t$ only through the combination $t \tilde t$.

\section{Results for refined Hilbert series } \label{Apprefres}

\subsection{$U(N_c)$ with $N_f=N_c+2$}

Here we prove Eq.~\eqref{eq:Nfp2res}. Starting from the general result \eqref{eq:Ungen}, we can proceed similarly as for $N_f=N_c+1$ (see Eq.~\eqref{eq:Ncplus1}):
\begin{eqnarray} \label{eq:Nfp2exp}
 g_{N_c+2,U(N_c)}(t,\tilde t,x,y) &=&\sum_{\la} (t\tilde t)^{|\la|} s_\la(\tilde x)  s_\la(\tilde y) - \sum_{\la:\ell(\la) = N_c+1} (t\tilde t)^{|\la|} s_\la(\tilde x)  s_\la(\tilde y) \\ \nonumber
&&- \sum_{\la:\ell(\la) = N_c+2} (t\tilde t)^{|\la|} s_\la(\tilde x)  s_\la(\tilde y) \ .
\end{eqnarray}
The first term can be immediately summed, and the last term can be treated as in the $N_f=N_c+1$ calculation, while
the middle term is more difficult to evaluate. It can however be summed by using a cofactor expansion. First,
we notice that as $\ell(\la)=N_c+1$, we can define a partition $\kappa$ such that $\kappa+1(N_c+1)=\lambda$, {\em{i.e.}},  $\kappa_i = \lambda_i-1$ for all $i=1,\ldots,N_c+1$, and $\kappa_{N_c+2}=0$.
Now, as $\kappa$ runs over all partitions with $\ell(\kappa)\le N_c+1$, $\lambda$ covers all partitions with $\ell(\la)=N_c+1$. By using the definition of $s_\la$,
\begin{equation}
\Delta(\tilde x) s_\lambda(\tilde x) =  \det\left(\tilde x_j^{\lambda_{N_f+1-i}+i-1}\right)_{i,j=1,\ldots,N_f} \ .
\end{equation}
Now the determinant may be processed by applying the cofactor expansion on its first row. Then the subdeterminant of the $k$th term in the expansion is
\begin{eqnarray} \label{eq:cofactors}
 \det\left(\!\left(\tilde x_j^{(k)}\right)^{\lambda_{N_f-i}+i}\right)_{i,j=1,\ldots,N_c+1}\!\!\! &=& \left[\prod_{i=1}^{N_c+1} \tilde x_i^{(k)}\right]^2 \det\left(\!\left(\tilde x_j^{(k)}\right)^{\kappa_{(N_c+1)-i+1}+i-1}\right)_{i,j=1,\ldots,N_c+1} \\ \nonumber
 &=& \frac{1}{\tilde x_k^2} \Delta(\tilde x^{(k)}) s_\kappa(\tilde x^{(k)}) \ ,
\end{eqnarray}
where $\tilde x^{(k)}$ is the sequence $(\tilde x_1,\tilde x_2,\ldots,\tilde x_{k-1},\tilde x_{k+1},\ldots,\tilde x_{N_f})$, {\emph{i.e}}, $\tilde x$ with the element $\tilde x_k$ removed. We find that
\begin{equation}
 s_\lambda(\tilde x) = \sum_{k=1}^{N_f}(-1)^k\frac{1}{\tilde x_k^2} \frac{\Delta(\tilde x^{(k)})}{\Delta(\tilde x)} s_\kappa(\tilde x^{(k)}) \ ; \qquad (\ell(\la)=N_c+1) \ .
\end{equation}
Inserting this in the second term of~\eqref{eq:Nfp2exp} we can sum over the partitions:

\begin{eqnarray}
 &&\sum_{\la:\ell(\la) = N_c+1} (t\tilde t)^{|\la|} s_\la(\tilde x)  s_\la(\tilde y)\\\nonumber &&= (t\tilde t)^{N_c+1}\sum_{k,\ell=1}^{N_f} (-1)^{k+\ell}\frac{1}{\tilde x_k^2} \frac{\Delta(\tilde x^{(k)})}{\Delta(\tilde x)}\frac{1}{\tilde y_\ell^2} \frac{\Delta(\tilde y^{(\ell)})}{\Delta(\tilde y)} \sum_\kappa (t\tilde t)^{|\kappa|} s_\kappa(\tilde x^{(k)}) s_\kappa(\tilde y^{(\ell)})\nonumber \\ \nonumber
&&=(t\tilde t)^{N_c+1} \sum_{k,\ell=1}^{N_f} (-1)^{k+\ell}\frac{1}{\tilde x_k^2} \frac{\Delta(\tilde x^{(k)})}{\Delta(\tilde x)}\frac{1}{\tilde y_\ell^2} \frac{\Delta(\tilde y^{(\ell)})}{\Delta(\tilde y)} \prod_{i,j=1}^{N_c+1} \frac{1}{1-t\tilde t \tilde x_i^{(k)}\tilde y_j^{(\ell)}} \ ,
\end{eqnarray}
where the sequence $y^{(\ell)}$ is defined analogously to $x^{(k)}$ above. Notice also that we were able to drop the constraint $\ell(\kappa)\le N_c+1$ since the modified sequences contain only $N_c+1$ variables. Combining with the other terms, and after some reorganization, we get
\begin{eqnarray} \label{eq:Nfp2resapp}
&&   g_{N_c+2,U(N_c)}(t,\tilde t,x,y) =  \prod_{i,j=1}^{N_f}\frac{1}{1-t\tilde t \tilde x_i \tilde y_j}\\ \nn
&&\times\Bigg[1-(t\tilde t)^{N_f} -(t\tilde t)^{N_c+1} \sum_{k,\ell=1}^{N_f} \frac{\prod_{i=1}^{N_f}(1-t\tilde t \tilde x_k \tilde y_i)\prod_{j=1}^{N_f}(1-t\tilde t \tilde x_j \tilde y_\ell)}{\tilde x_k^2 \tilde y_\ell^2 (1-t\tilde t \tilde x_k \tilde y_\ell)\prod_{i\ne k} (\tilde x_i-\tilde x_k) \prod_{j\ne \ell} (\tilde y_j-\tilde y_\ell)} \Bigg] \ .
\end{eqnarray}

\subsection{$SU(N_c)$ with $N_f=N_c+1$}

Here we prove Eq.~\eqref{eq:SUNcp1res}. The proof is quite similar to the one above for $N_f=N_c+2$ with the $U(N_c)$ gauge group. We start from Eq.~\eqref{eq:IkNfgNc} taking $k \ge 0$:
\begin{equation}
I_k = t^{kN_c} \sum_{\kappa:\ell(\kappa)\le N_c}  t^{|\kappa|} s_{\kappa+k(N_c)}(\tilde x) \tilde  t^{|\kappa|} s_\kappa(\tilde y) \ .
\end{equation}
Proceeding as above (see Eq.~\eqref{eq:cofactors}),
\begin{equation}
 s_{\kappa+k(N_c)}(\tilde x)  =  \sum_{m=1}^{N_f}(-1)^m\frac{1}{\tilde x_m^{k+1}} \frac{\Delta(\tilde x^{(m)})}{\Delta(\tilde x)} s_\kappa(\tilde x^{(m)}) \ .
\end{equation}
Hence we get
\begin{eqnarray}
 I_k &=& t^{kN_c} \sum_{m,n=1}^{N_f}(-1)^{m+n}\frac{1}{\tilde x_m^{k+1}} \frac{\Delta(\tilde x^{(m)})}{\Delta(\tilde x)}\frac{1}{\tilde y_n} \frac{\Delta(\tilde y^{(n)})}{\Delta(\tilde y)} \sum_\kappa(t\tilde t)^{|\kappa|} s_\kappa(x^{(m)}) s_\kappa(y^{(n)}) \\ \nonumber
&=&t^{kN_c} \sum_{m,n=1}^{N_f}(-1)^{m+n}\frac{1}{\tilde x_m^{k+1}} \frac{\Delta(\tilde x^{(m)})}{\Delta(\tilde x)}\frac{1}{\tilde y_n} \frac{\Delta(\tilde y^{(n)})}{\Delta(\tilde y)}\prod_{i,j=1}^{N_c} \frac{1}{1-t\tilde t \tilde x_i^{(m)}\tilde y_j^{(n)}} \ ; \qquad (k \ge 0) \ .
\end{eqnarray}
A similar calculation for $k<0$ yields
\begin{equation}
 I_k = \tilde t^{|k|N_c} \sum_{m,n=1}^{N_f}(-1)^{m+n}\frac{1}{\tilde x_m} \frac{\Delta(\tilde x^{(m)})}{\Delta(\tilde x)}\frac{1}{\tilde y_n^{|k|+1}} \frac{\Delta(\tilde y^{(n)})}{\Delta(\tilde y)}\prod_{i,j=1}^{N_c} \frac{1}{1-t\tilde t \tilde x_i^{(m)}\tilde y_j^{(n)}} \ ; \qquad (k < 0) \ .
\end{equation}
Inserting the result in the definition~\eqref{eq:Ik} the sum over $k$ can be done, giving
\begin{eqnarray}
 && g_{N_c+1,SU(N_c)}(t,\tilde t,x,y) \\\nonumber
&& =\ \sum_{m,n=1}^{N_f}(-1)^{m+n}\left(\frac{1}{1-t^{N_c}/\tilde x_m}+\frac{1}{1-\tilde t^{N_c}/\tilde y_n}-1\right)\frac{1}{\tilde x_m^{1}} \frac{\Delta(\tilde x^{(m)})}{\Delta(\tilde x)}\frac{1}{\tilde y_n^{|k|+1}} \frac{\Delta(\tilde y^{(n)})}{\Delta(\tilde y)} \\\nonumber
&&\ \ \times \prod_{i,j=1}^{N_c} \frac{1}{1-t\tilde t \tilde x_i^{(m)}\tilde y_j^{(n)}} \\ \nonumber
&&= \ \prod_{i,j=1}^{N_f} \frac{1}{1-t\tilde t \tilde x_i \tilde y_j} \\ \nonumber
&&\ \  \times\! \sum_{k,\ell=1}^{N_f}\! \frac{(1-(t\tilde t)^{N_c}/(\tilde x_k\tilde y_\ell))\prod_{i=1}^{N_f}(1-t\tilde t \tilde x_k \tilde y_i)\prod_{j=1}^{N_f}(1-t\tilde t \tilde x_j \tilde y_\ell)}{\tilde x_k \tilde y_\ell(1-t^{N_c}/\tilde x_k)(1-\tilde t^{N_c}/\tilde y_\ell) (1-t\tilde t \tilde x_k \tilde y_\ell)\prod_{i\ne k} (\tilde x_i-\tilde x_k) \prod_{j\ne \ell} (\tilde y_j-\tilde y_\ell)} \ .
\end{eqnarray}

\section{Asymptotics of the unrefined Hilbert series} \label{Appdasympt}

In this Appendix we analyze the asymptotics of the unrefined Hilbert series and the singularity structures of the consequent sums.
First, we recall that
\begin{eqnarray}
 g_{N_f,U(N_c)}(t,\tilde t) &=& \sum_{s=0}^\infty(t \tilde t)^s G(s,0) \\
  g_{N_f,SU(N_c)}(t,\tilde t) &=& \sum_{k=0}^\infty \sum_{s=0}^\infty t^{k M} (t\tilde t)^s G(s,k) + \sum_{k=-\infty}^{-1} \sum_{s=0}^\infty \tilde t^{|k| M} (t\tilde t)^s   G(s,|k|) \ , \label{eq:SUNunrefapp}
\end{eqnarray}
where
\begin{eqnarray} \label{eq:Gdefapp}
 G(s,k) &\equiv& \sum_{\la: |\la|=s,\ \ell(\la) \le M} d_{\la+k(N_c)} d_\la \\
 M &\equiv& \min\left(N_c,N_f\right) \ .
\end{eqnarray}
Therefore, we need to find the asymptotics of $G(s,k)$ for large $s$ and $k$. We do this by a ``brute force'' calculation using the standard formula
\begin{equation}
 d_\la = \prod_{m\in \la} \frac{N_f+i(m)-j(m)}{h(m)} \ ,
\end{equation}
where the box $m$ of the Young diagram $\la$ lies on the $i(m)$th row and the $j(m)$th column, and $h(m)$ is the hook length associated to $m$:
\begin{equation}
 h(m) = \la_{j(m)} - i(m)+\la'_{i(m)}-j(m)+1
\end{equation}
with $\la'$ being the conjugate partition, {\em{i.e.}}, $\la'_i$ is the length of the $i$th row of $\la$.

Let us abbreviate $N_f=N$. We shall divide the diagram $\la$ into $M$ rectangular blocks, denoted by $\la^{(K)}$, and defined by
\begin{equation}
 m \in \la^{(K)} \Leftrightarrow \la'_{i(m)} = K \ , \qquad K=1,\ldots,M\ ,
\end{equation}
{\em{i.e.}}, the $K$th block consists of rows that have the length $K$.
Then we have
\begin{eqnarray} \label{eq:dimdecomp}
 d_\la &=& \prod_{K=1}^M d_{\la^{(K)}} \\
 d_{\la^{(K)}} &\equiv&  \prod_{m\in \la^{(K)}} \frac{N+i(m)-j(m)}{h(m)} \ . 
\end{eqnarray}
Since the blocks are rectangular, it is straightforward to write $d_{\la^{(K)}}$ in terms of Euler gamma functions. At this point, let us also reintroduce the variable $k$.\footnote{We shall consider the partition $\la+k(M)$, {\em{i.e.}}, not $\la+k(N_c)$. Whenever $N_c \ne M$ ({\em{i.e.}}, $M=N_f<N_c$) only the $k=0$ term contributes in the sums above, and $s_{\la+k(N_c)}(\tilde x)=0$.}
We get
\begin{eqnarray}
  d_{(\la+k(M))^{(K)}} &=& \prod_{i=\la_{K+1}+k+1}^{\la_K+k} \prod_{j=1}^K \frac{N+i-j}{\la_j+k-i+K-j+1} \\\nonumber
&=&  \prod_{j=1}^K \frac{\Gamma(N+\la_{K}+k-j+1)}{\Gamma(N+\la_{K+1}+k-j+1)} \frac{\Gamma(\la_j-\la_K+K-j+1)}{\Gamma(\la_j-\la_{K+1}+K-j+1)}
\end{eqnarray}
for $1\le K <M$ and
\begin{equation}
   d_{(\la+k(M))^{(M)}} =  \prod_{j=1}^M  \frac{\Gamma(N+\la_{M}+k-j+1)}{\Gamma(N-j+1)} \frac{\Gamma(\la_j-\la_M+M-j+1)}{\Gamma(\la_j+k+M-j+1)} \ .
\end{equation}
In order to find $G(s,k)$ at large $s=|\la|= \sum_i \la_i$ and $k$ we need to analyze these expressions in the limit where $\la_i \to \infty$ with their ratios fixed, and also $k \to \infty$. We shall not take $s/k$ fixed, but rather consider limits where either of them is taken to infinity independently. The analysis can be done by using the asymptotics of the gamma functions, {\em{i.e.}}, the Stirling formula
\begin{equation}
 \log \Gamma(\la+x) = \la\log\la - \la +\left(x-\frac{1}{2}\right)\log\la +\frac{1}{2}\log(2\pi) + \mathcal{O}\left(\frac{1}{\la}\right) \ ,
\end{equation}
where we neglect the constant term, since we are not trying to calculate the proportionality constant of the final scaling results. We also notice immediately that the term $\sim -\la$ does not contribute due to cancellations. Denoting $f(x) =x\log x$, we get
\begin{eqnarray} \label{eq:dKasymp}
&& d_{(\la+k(M))^{(K)}}\\\nonumber
&& \sim \  \exp\bigg[\sum_{j=1}^Kf(\la_K\!+\!k) - \sum_{j=1}^Kf(\la_{K+1}\!+\!k) + \sum_{j=1}^{K-1} f(\la_j\!-\!\la_K)-\sum_{j=1}^{K} f(\la_j\!-\!\la_{K+1}) \bigg] \\\nonumber
&&\  \ \times \exp\bigg[\sum_{j=1}^K(N-j+1/2)\log(\la_K\!+\!k)-\sum_{j=1}^K(N-j+1/2)\log(\la_{K+1}\!+\!k)\bigg] \\\nonumber
&&\ \ \times \exp\bigg[\sum_{j=1}^{K-1}(K-j+1/2)\log(\la_j\!-\!\la_K)-\sum_{j=1}^K(K-j+1/2)\log(\la_{j}\!-\!\la_{K+1})\bigg]
\end{eqnarray}
for $1\le K <M$. We left out all factors that are finite as $s \to\infty$ (with all $\la_i/\la_j$ fixed)  or $k\to \infty$. Notice also that the sum of the log-terms (last two rows in~\eqref{eq:dKasymp}) is finite in both limits if the $j=K$ term in the last sum is removed. Therefore, we shall only keep this divergent term below. For $K=M$ we find
\begin{eqnarray}
&& d_{(\la+k(M))^{(M)}}\\\nonumber
&& \sim\  \exp\bigg[\sum_{j=1}^Mf(\la_M\!+\!k)  + \sum_{j=1}^{M-1} f(\la_j\!-\!\la_M)-\sum_{j=1}^{M} f(\la_j\!+\!k) \bigg] \\\nonumber
&&\ \  \times \exp\bigg[\sum_{j=1}^M(N-j+1/2)\log(\la_M\!+\!k)\bigg] \\\nonumber
&&\ \ \times \exp\bigg[\sum_{j=1}^{M-1}(M-j+1/2)\log(\la_j\!-\!\la_M)-\sum_{j=1}^M(M-j+1/2)\log(\la_{j}\!+\!k)\bigg] \ .
\end{eqnarray}
Let us then insert the results in the formula~\eqref{eq:dimdecomp}. It is not difficult to check that all terms involving $f$ cancel. We are left with
\begin{eqnarray}
d_{\la+k(M)}
& \sim & \exp\bigg[-\frac{1}{2}\sum_{K=1}^{M-1}\log(\la_K\!-\!\la_{K+1})  + \sum_{j=1}^M(N-j+1/2)\log(\la_M\!+\!k) \\\nonumber
&& + \sum_{j=1}^{M-1}(M-j+1/2)\log(\la_j\!-\!\la_M)-\sum_{j=1}^M(M-j+1/2)\log(\la_{j}\!+\!k)\bigg] \ .
\end{eqnarray}
To continue, we write an expression which correctly interpolates between the leading asymptotics as either $s=|\la| \to \infty$ or $k \to \infty$ first.\footnote{A more precise way would be to simply treat the limits separately. Then we would not need to introduce the $\mathcal{O}(1)$ functions $C$ which depend in a complicated way on $\la_i/s$ and will cancel in all results.} We get formally
\begin{eqnarray} \label{eq:dlakscal}
d_{\la+k(M)} 
& \sim& \exp\bigg[-\frac{1}{2}\sum_{K=1}^{M-1}\log s  + \sum_{j=1}^M(N-j+1/2)\log(s\!+\!C k) \\\nonumber
&& + \sum_{j=1}^{M-1}(M-j+1/2)\log s-\sum_{j=1}^M(M-j+1/2)\log(s\!+\!C k)\bigg] \\\nonumber
& \sim & (s\!+\!C k)^{M(N-M)} s^{M(M-1)/2} \ ,
\end{eqnarray}
where $C$ can be treated as a constant.
The remaining task is to sum over the partitions in the definition~\eqref{eq:Gdefapp}.
As the scaling result~\eqref{eq:dlakscal} is independent of $\la_i$, this amounts to calculating
the number of partitions that enter the sum. We sum over $\la_1 \ge \la_2 \ge \cdots \ge \la_M$ with one
constraint $s=\sum_i\la_i$, so the number of terms is $\sim s^{M-1}$. Therefore,\footnote{The result~\eqref{eq:Gscaling}
covers all nontrivial cases, {\em{i.e.}}, it holds for $k=0$ or $N_f \ge N_c$. For $k\ne 0$ and $N_f<N_c$ the result is not
applicable since we used the formula for $d_{\la+k(M)}$ instead of $d_{\la+k(N_c)}$, but in this case actually $G(s,k)=0$. }
\begin{eqnarray} \label{eq:Gscaling}
 G(s,k) &\sim&  s^{M-1} d_{\la+k(M)} d_{\la+0(M)} \\\nonumber
&=&  s^{M-1} (s\!+\!C k)^{M(N-M)} s^{M(M-1)/2} s^{M(N-M)} s^{M(M-1)/2} \\\nonumber
&=&  (s\!+\!C k)^{M(N-M)} s^{MN-1} \ .
\end{eqnarray}

The main results are given in the text where we assume that $N_f \ge N_c$ so that $M=N_c$ (and $N=N_f$). However, the analysis also works
when $N_f<N_c$ and $M=N_f$, even though only the $k=0$ case appears in the sums~\eqref{eq:SUNunrefapp} in this region. Indeed, we find
\begin{equation}
 G(s,0) \sim  s^{N_f^2-1}
\end{equation}
and
\begin{equation}
 g_{N_f,U(N_c)}(t,\tilde t)= g_{N_f,SU(N_c)}(t,\tilde t) \sim \sum_{s=0}^\infty(t \tilde t)^s  s^{N_f^2-1} \sim \frac{1}{(1-t \tilde t)^{N_f^2}}
\end{equation}
for $N_f<N_c$ as $t\tilde t \to 1$, which is the expected result (compare with (\ref{eq:unrefUN}) with $\tilde x_i=\tilde y_j=1$ for all $i,j$).

\section{Saddle point analysis of the GCBO result} \label{AppGCBO}

In this Appendix we shall calculate the leading asymptotics of the results~\eqref{eq:GCBO1} and~\eqref{eq:GCBO2} in the Veneziano limit.
We start from the standard integral formula
\beq \label{eq:Freporg}
 _2F_1(a,b;c;z)= \frac{\Gamma(c)}{\Gamma(b)\Gamma(c-b)} \int_0^1 ds s^{b-1}(1-s)^{c-b-1}(1-zs)^{-a} \ .
\eeq
However, in our case $b>c$ so the integral will not converge at $t=1$. We then regulate the $t=1$ singularity in a standard manner: we deform the
integration contour and write
\beq \label{eq:Frep}
 _2F_1(a,b;c;z)= \frac{\Gamma(c)}{\Gamma(b)\Gamma(c-b)} \frac{1}{1-e^{2\pi i(c-b)}} \int_\mC ds s^{b-1}(1-s)^{c-b-1}(1-zs)^{-a} \ ,
\eeq
where the integration contour starts from the origin, circles around $s=1$ in the counterclockwise direction, and returns to $s=0$ via a different
branch of the integrand. We can contract $\mC$ to two pieces of a real line and a circle around $s=1$ having radius $\epsilon$.
In the domain $\re (c)>\re (b)$ the contribution on the circle vanishes as $\epsilon \to 0$ and we recover the original
representation~\eq{Freporg}, which proves the formula~\eq{Frep}.

\begin{figure}[ht]
\centering
\includegraphics[width=0.4\textwidth]{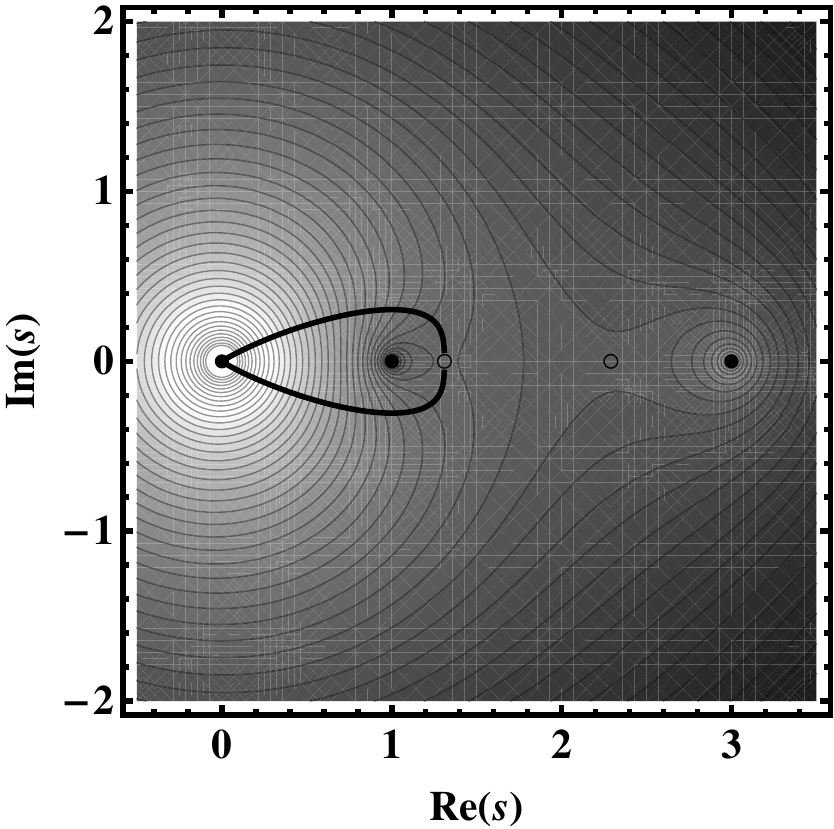}\hspace{1cm}%
\includegraphics[width=0.4\textwidth]{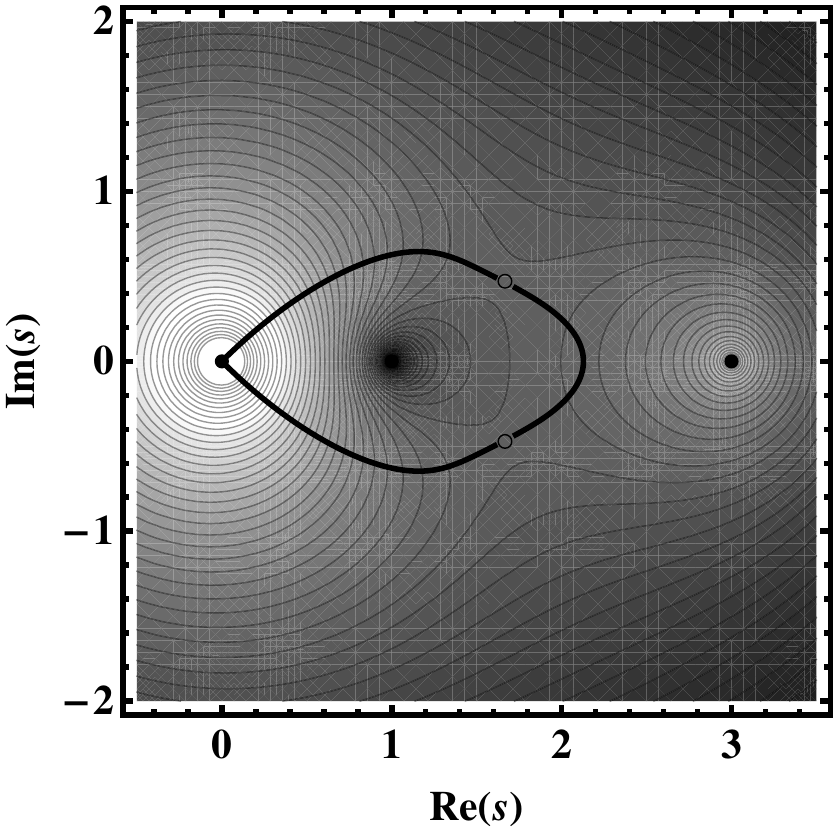}
\caption{\label{fig1}An illustration for the saddle point method on the complex $s$-plane. Left: saddle points at real values of $s$. Right: saddle points at complex $s$. The contours in the background are at constant values of $\mathrm{Re}\, F(s)$ with lighter shades denoting smaller values. The filled circles are singularities of $F(s)$, while the open circles are the saddle points. The integration contour $\mathcal{C}$ is the thick curve that starts at $s=0$, goes around the singularity at $s=1$ through one (left) or both (right) of the saddle points, and returns to $s=0$.}
\end{figure}

Now the integral representation converges in the domain of interest to us and we can proceed with the saddle point analysis. Let us first rewrite the representation as
\beqa
_2F_1(a,b;c;z) &=& \frac{\Gamma(c)\Gamma(1+b-c)}{\Gamma(b)} I \\
I &=& \inv{2\pi i} \int_\mC ds s^{b-1}(s-1)^{c-b-1}(1-zs)^{-a} \ .
\eeqa
We will restrict to the leading result (next-to-leading corrections will be considered below)
so we
take $a=-(r-1)N_c$, $b-1=rN_c$ and $c-b-1=-(r-1)N_c$ which covers both functions in~\eq{Fdefs}. Also recall that $z=t\tilde t$ lies between zero and one. The crucial function is the logarithm of the integrand:
\beq
 F(s) = \left[r\log s-(r-1)\log(s-1)+(r-1)\log(1-sz)\right] \ ,
\eeq
where we left out the common factor $N_c$.
Saddle points are found at the zeroes of
\beq
  F'(s) =\frac{r}{s}-\frac{r-1}{s-1}-\frac{z(r-1)}{1-sz} = \frac{-r z s^2 + s(1+2rz-z)-r}{s(s-1)(1-sz)}
\eeq
and are given by
\beq
 s_{\pm} = 1 +\frac{1-z\pm\sqrt{\left(1-z\right)\left[1-(2r-1)^2z \right]}}{2rz} \ .
\eeq
We observe that there are two possibilities, separated by the critical value $z=t\tilde t =1/(2r-1)^2$. We immediately notice that this
matches with the critical value of $t$ where the gap appears in the Coulomb gas approach.
Let us analyze the two cases separately.
\begin{itemize}
\item For $z=t \tilde t< 1/(2r-1)^2$ there are two real roots. It is easy to see that they lie between $s=1$ and $s=1/z$.
Since $F'(s) \to -\infty$ when $s$ approaches one from above,  $F''(s_-)>0$ and $F''(s_+)<0$.
We choose $\mC$ such that it passes through $s=s_-$ perpendicularly to the real axis.
The leading behavior is then
\beq
 I \sim e^{N_c F(s_-)} \ ,
\eeq
where the proportionality constant is real and positive. We did not find a simple formula for $F(s_-)$. Using this result we find that
\beq \label{eq:corrtermsc}
 {r N_c\choose N_c} \widehat{\mathcal{F}}(N_c,r,z) z^{N_c/2} \sim {r N_c\choose N_c+1} \mathcal{F}(N_c,r,z) z^{N_c/2} \sim e^{N_c \left[F(s_-)+\inv{2}\log z\right]} \ ,
\eeq
where the binomial coefficients and gamma functions cancel at leading order.
It is straightforward to show that the combination $F(s_-)+(\log z)/2$ is negative for $z<1/(2r-1)^2$ (and approaches zero as $z \to 1/(2r-1)^2$). Therefore, the corrections are exponentially suppressed in $N_c$ in the $U(N_c)$ case. For $SU(N_c)$ gauge symmetry, corrections are suppressed when $t=\tilde t$. For $t \ne \tilde t$ there are exponentially growing corrections for $z$ sufficiently close to $1/(2r-1)^2$, as the result for $\widehat{\mathcal{F}}$ in Eq.~\eqref{eq:corrtermsc} will be multiplied by
\begin{equation}
 \left(\frac{t}{\tilde t}\right)^{N_c/2}+\left(\frac{\tilde t}{t}\right)^{N_c/2} \ .
\end{equation}

\item For $z=t \tilde t> 1/(2r-1)^2$ there are two complex roots:
\beq
 \hat s_{\pm} = 1 +\frac{1-z\pm i \sqrt{\left(1-z\right)\left[(2r-1)^2z-1 \right]}}{2rz} \ .
\eeq
Let us denote $F(\hat s_\pm) = A \pm i B$. A straightforward calculation yields $A=-(\log z)/2$. We find that all exponential factors cancel:
\beq
 {r N_c\choose N_c} \widehat{\mathcal{F}}(N_c,r,z) z^{N_c/2} \sim {r N_c\choose N_c+1} \mathcal{F}(N_c,r,z) z^{N_c/2} \sim \cos(N_c B+\phi) \ .
\eeq
The vividly oscillating $\mathcal{O}(1)$ result suggests that the approximation of Eqs.~\eqref{eq:GCBO1} and~\eqref{eq:GCBO2} is breaking down.

\end{itemize}

Finally, let us consider higher order corrections to the saddle point results. The exact form of the logarithm of the integrand in the integral representation of the hypergeometric function is
\beq
 \tilde F(s) = N_c F(s) + G(s) \ ,
\eeq
where $G(s)$ differs for the two expressions of Eq.~\eqref{eq:Fdefs}: the parameters  $a = -(r-1)N_c+1$ and $b=rN_c$ of the hypergeometric functions are the same in both cases, but $c$ is different. Therefore, we define
\beq
 G(s) = \left\{\begin{array}{rcl}
     \displaystyle   \log(s-1)-\log(s)-\log(1-zs) \equiv G_1(s)  \ ; \mbox{\qquad} &  c= N_c+2 \ \phantom{.}\\
   \displaystyle              -\log(s)-\log(1-zs) \equiv G_2(s) \ ; \mbox{\qquad} &  c= N_c+1 \ .
     \end{array}\right.
\eeq
We shall denote 
the ``corrected'' zeroes of $\tilde F'(s)$ as $\tilde s_\pm$. The full Gaussian contribution to the saddle point result,
\begin{eqnarray}
 I &\simeq& \frac{1}{\sqrt{2 \pi\tilde  F''(\tilde s_-)}} e^{\tilde F(\tilde s_-)} \ \ \qquad \mathrm{for\ real\ roots} \\
 I &\simeq& \frac{1}{\sqrt{2 \pi\tilde  F''(\tilde s_+)}} e^{\tilde F(\tilde s_+)} +\frac{1}{\sqrt{2 \pi F''(\tilde s_-)}} e^{\tilde F(\tilde s_-)}\\\nonumber
 &=& 2 \mathrm{Re}\left[\frac{1}{\sqrt{2 \pi\tilde  F''(\tilde s_+)}} e^{\tilde F(\tilde s_+)}\right] \ \  \qquad \mathrm{for\ complex\ roots} \ ,
\end{eqnarray}
is accurate up to $\mathcal{O}(1/N_c)$ corrections. Expanding in $1/N_c$ we find, after some calculations,
\begin{eqnarray}
  I &=&  \frac{1}{\sqrt{2 \pi N_c F''(s_-)}}e^{N_c F(s_-) +G(s_-)} \left[1+\mathcal{O}\left(\frac{1}{N_c}\right)\right] \ ; \qquad t\tilde t <1/(2r-1)^2 \\ \nonumber
  I &=&  2 \mathrm{Re}\left[\frac{1}{\sqrt{2 \pi N_c F''(\hat s_+)}}e^{N_c F(\hat s_+) +G(\hat s_+)}\right] \left[1+\mathcal{O}\left(\frac{1}{N_c}\right)\right]   \ ; \qquad t\tilde t > 1/(2r-1)^2 \ .
\end{eqnarray}
The correction terms become
\begin{eqnarray}
 {r N_c\choose N_c+1} \mathcal{F}(N_c,r,z) z^{N_c/2} &=&  \frac{1}{\sqrt{2 \pi N_c F''(s_-)}}\frac{r}{r-1}  \\ \nonumber
 &&\times \exp\left[N_c F(s_-)+\frac{N_c}{2}\log z +G_1(s_-)\right] \\
 {r N_c\choose N_c} \widehat{\mathcal{F}}(N_c,r,z) z^{N_c/2} &=&  \frac{1}{\sqrt{2 \pi N_c F''(s_-)}}\frac{r}{r-1} \\ \nonumber
&& \times \exp\left[N_c F(s_-)+\frac{N_c}{2}\log z +G_2(s_-)\right]
\end{eqnarray}
in the region of real saddle points, and
\begin{eqnarray}
 {r N_c\choose N_c+1} \mathcal{F}(N_c,r,z) z^{N_c/2} &=&  \frac{\sqrt{2}}{\sqrt{\pi N_c |F''(\hat s_+)|}}\frac{r}{r-1}\cos(N_cB + \phi_1)  \\\nn
 {r N_c\choose N_c} \widehat{\mathcal{F}}(N_c,r,z) z^{N_c/2} &=&  \frac{\sqrt{2}}{\sqrt{ \pi N_c |F''(\hat s_+)|}}\frac{r}{r-1}\sqrt{\frac{r z}{(r-1)(1-z)}}\cos(N_cB + \phi_2)
\end{eqnarray}
in the region of complex saddle points, up to $1/N_c$ corrections. Here $B=\mathrm{Im} F(\hat s_+)$ as above,
\begin{equation}
 \phi_i = -\frac{1}{2}\arg F''(\hat s_+) + \mathrm{Im}\, G_i(\hat s_+) \ ,
\end{equation}
and we used the results
\begin{equation}
 \mathrm{Re}\,G_1(\hat s_+) = 0\ ; \qquad \mathrm{Re}\,G_2(\hat s_+) = \frac{1}{2}\log \frac{r z}{(r-1)(1-z)} \ .
\end{equation}

\end{document}